%% file: main.tex
 \documentclass[sigconf]{acmart}

\usepackage{amsmath,amssymb,amsfonts}
\usepackage{graphicx}
\usepackage{textcomp}
\usepackage{tabularx}
\usepackage{enumitem}
\usepackage{todonotes}
\usepackage{framed}
\usepackage{multirow}
\usepackage{pifont}
\usepackage{lipsum}
\usepackage{etoolbox}
\usepackage{algorithm}
\usepackage[noend]{algpseudocode}
\usepackage{listings}
\usepackage{tabto}
\usepackage{verbatim}
\usepackage{lstlinebgrd}
\usepackage{mwe}

\newcommand{\squishlist}{
 \begin{list}{$\bullet$}
  { \setlength{\itemsep}{0pt}
     \setlength{\parsep}{3pt}
     \setlength{\topsep}{3pt}
     \setlength{\partopsep}{0pt}
     \setlength{\leftmargin}{1.5em}
     \setlength{\labelwidth}{1em}
     \setlength{\labelsep}{0.5em} } }

\newcommand{\squishend}{
  \end{list} 
}

\algnewcommand{\IIf}[1]{\State\algorithmicif\ #1\ \algorithmicthen}
\algnewcommand{\EElse}[1]{\State\algorithmicelse\ #1\ }
\algnewcommand{\EndIIf}{\unskip}
\algrenewcommand\algorithmicindent{1.2em}%

\newcommand{\CC}{C\nolinebreak\hspace{-.05em}\raisebox{.01ex}{\normalfont  +}\nolinebreak\hspace{-.10em}\raisebox{.01ex}{\normalfont +}}

\definecolor{dkgreen}{rgb}{0,0.3,0}
\definecolor{gray}{rgb}{0.5,0.5,0.5}
\definecolor{mauve}{rgb}{0.58,0,0.82}
\definecolor{dkpink}{RGB}{167,29,93}
\definecolor{dkblue}{RGB}{24,54,145}
\definecolor{greenCode}{RGB}{153, 213, 202}
\definecolor{blueCode}{RGB}{131, 187, 229}
\definecolor{yellowCode}{RGB}{252,196,56}

\definecolor{listinggray}{gray}{0.9}
\definecolor{lbcolor}{rgb}{0.9,0.9,0.9}

\lstdefinelanguage{XML}
{
	morestring=[s][\color{dkpink}]{"}{"},
	morestring=[s][\color{black}]{>}{<},
	morecomment=[s]{<?}{?>},
	morecomment=[s][\color{dkgreen}]{<!--}{-->},
	stringstyle=\color{black},
	identifierstyle=\color{dkblue},
	keywordstyle=\color{red},
	morekeywords={xmlns,xsi,noNamespaceSchemaLocation,type,id,x,y,source,target,version,tool,transRef,roleRef,objective,eventually}%
}

\input{EpsilonFormat.tex}

\newcommand{\code}[1]{{\texttt{#1}}}

\AtBeginDocument{%
  \providecommand\BibTeX{{%
    \normalfont B\kern-0.5em{\scshape i\kern-0.25em b}\kern-0.8em\TeX}}}

\setcopyright{acmcopyright}
\copyrightyear{2020}
\acmYear{2020}
\acmDOI{10.1145/1122445.1122456}

\acmConference[Woodstock '18]{Woodstock '18: ACM Symposium on Neural
  Gaze Detection}{June 03--05, 2018}{Woodstock, NY}
\acmBooktitle{Woodstock '18: ACM Symposium on Neural Gaze Detection,
  June 03--05, 2018, Woodstock, NY}
\acmPrice{15.00}
\acmISBN{978-1-4503-XXXX-X/18/06}

\newcommand{\approach}{RoboSMi}

\begin{document}

\title[Supporting Robotic Software Migration Using Static Analysis and Model-Driven Engineering]{
Supporting Robotic Software Migration \\Using Static Analysis and Model-Driven Engineering}

\author{Sophie Wood}
\affiliation{%
  \institution{University of York, York, UK}
}
\email{soph.wood52@gmail.com}

\author{Nicholas Matragkas}
\affiliation{%
  \institution{University of York, York, UK}
}
\email{nicholas.matragkas@york.ac.uk}

\author{Dimitris Kolovos}
\affiliation{%
  \institution{University of York, York, UK}
}
\email{dimitris.kolovos@york.ac.uk}

\author{Richard Paige}
\affiliation{%
  \institution{McMaster University, Canada}
}
\email{paigeri@mcmaster.ca}

\author{Simos Gerasimou}
\affiliation{%
  \institution{University of York, York, UK}
}
\email{simos.gerasimou@york.ac.uk}

\renewcommand{\shortauthors}{Wood et al.}



\begin{abstract}

\noindent
The wide use of robotic systems 
contributed to developing robotic software highly coupled to the hardware platform running the robotic system. Due to increased maintenance cost 
or changing business priorities, the robotic hardware is infrequently upgraded, thus increasing the risk for technology stagnation. 
Reducing this risk entails migrating the system and its software to a new hardware platform. Conventional software engineering practices such as complete re-development and code-based migration, albeit useful in mitigating these obsolescence issues, 
they are time-consuming and overly expensive.  Our RoboSMi model-driven approach supports the migration of the software controlling a robotic system between hardware platforms. First, RoboSMi executes static analysis on the robotic software of the source hardware platform to identify platform-dependent and platform-agnostic software constructs. By analysing a model that expresses the architecture of robotic components on the target platform, RoboSMi establishes the hardware configuration of those components and suggests software libraries for each component whose execution will enable the robotic software to control the components. 
Finally, RoboSMi through code-generation produces software for the target platform and indicates areas that require manual intervention by robotic engineers to complete the migration. We evaluate the applicability of \approach\ and analyse the level of automation and performance provided from its use by migrating 
two robotic systems deployed for an environmental monitoring and a line following mission from a Propeller Activity Board to an Arduino Uno.
\vspace*{-1mm}
\end{abstract}

\maketitle

\input{sections/s1-introduction}
\input{sections/s2-problem}

\input{sections/s3-methodology}
\input{sections/s4-implementation}

\input{sections/s5-evaluation}

\input{sections/s6-relatedWork}
\input{sections/s7-conclusion}

\bibliographystyle{ACM-Reference-Format}
\bibliography{main}
\end{document}

%% file: EpsilonFormat.tex
\lstdefinelanguage{Emfatic}{
	alsoletter={@},
	keywords={[1]attr,class,datatype,enum,import,interface,mapentry,op,package,ref,val},
	keywordstyle={[1]\color{blue}\bfseries\underbar},
	morekeywords={[2]abstract, derived, extends, super, false, id, ordered,
		readonly, resolve, throws, transient, true, unique, unsettable, void,
		volatile},
	keywordstyle={[2]\color{MyDarkBlue}}, 
	morekeywords={@namespage},
	alsodigit={[]*()},
	alsoother={;/:.}, 
	morekeywords={[3]OrderedSet,Sequence,Set,Bag,EBoolean, EByte, EChar, EDouble, EFloat, EInt, 
		ELong, EShort, EBooleanObject, EByteObject, ECharacterObject, 
		EDoubleObject, EFloatObject, EIntegerObject, ELongObject, EShortObject, EDate, EString, EJavaClass, 
		boolean, byte, char, double, float, int, long, short, Boolean, 
		Byte, Character, Double, Float, Integer, Long, Short, Date, 
		String, Object, Class, EObject, EClass, 
		EDate, EBigInteger, EBigDecimal, EResource, EResourceSet, 
		EEnumerator, EEList, ETreeIterator, EJavaObject}, 
	keywordstyle={[3]\color{dark-green}}, 
	string=[d]",
	sensitive=true,
	morecomment=[l]{//},
	morecomment=[s]{/*}{*/},
	keywordstyle=\color{blue}\bfseries,
	identifierstyle=, 
	commentstyle=\color{light-gray}, %
	stringstyle=\textit, %
	columns=flexible
}[keywords, comments, strings]  

\lstdefinelanguage{EOL}{
	morekeywords={delete,import,for,while,in,and,or,self,operation,return,def,var,throw,if,new,else,transaction,abort,break,continue,assert,assertError,not},
	sensitive=true,
	morecomment=[l]{//},
	morestring=[b]',
	showstringspaces=false
}

\lstdefinelanguage{EVL}{
	morekeywords={guard,delete,import,for,while,in,and,or,self,operation,return,def,var,throw,if,new,else,transaction,abort,break,continue,assert,assertError,not,context,critique,constraint,check,message},
	sensitive=true,
	morecomment=[l]{//},
	morestring=[b]',
	showstringspaces=false
}

\lstdefinelanguage{EGL}{
	morekeywords={delete,import,for,while,in,and,or,self,operation,return,def,var,throw,if,for,new,else,transaction,abort,break,continue,assert,assertError,not,[\%, \%],[\%=},
	sensitive=true,
	morecomment=[l]{//},
	morestring=[b]',
	showstringspaces=true
}

\lstdefinelanguage{ETL}{
	morekeywords={delete,import,for,while,in,and,or,self,operation,return,def,var,throw,if,new,else,transaction,abort,break,continue,assert,assertError,not,rule,transform,to},
	sensitive=true,
	morecomment=[l]{//},
	morestring=[b]',
	showstringspaces=false
}

\lstdefinelanguage{CSS}{
	keywords={color,background-image:,margin,padding,font,weight,display,position,top,left,right,bottom,list,style,border,size,white,space,min,width, transition:, transform:, transition-property, transition-duration, transition-timing-function,fontColor}, 
	sensitive=true,
	morecomment=[l]{//},
	morecomment=[s]{/*}{*/},
	morestring=[b]',
	morestring=[b]",
	alsoletter={:},
	alsodigit={-}
}

\lstdefinelanguage{OCL}{
	morekeywords={delete,import,for,while,in,and,or,self,operation,return,def,var,throw,if,new,else,transaction,abort,break,continue,assert,assertError,not,true, false, endif, in},
	sensitive=true,
	morecomment=[l]{//},
	morestring=[b]',
	showstringspaces=false
}

%% file: sections/s1-introduction.tex

\section{Introduction}
\label{sec:introduction}
\noindent
Robotic systems are increasingly used in various application domains ranging from transportation~\cite{farinelli2017advanced} and healthcare~\cite{hawes2017strands} to agriculture~\cite{bergerman2016robotics} and warehouse management~\cite{wang2017ubiquitous}. Driven by recent technological advancements, these systems provide sophisticated functionality, increased efficiency and automation by assisting or, when possible, replacing human operators in repetitive, laborious or potentially dangerous tasks~\cite{siciliano2016springer}. For instance, mobile robots deployed within a warehouse facility can perform automatic inventory checks, 
thus reducing the need for manual inventory counts. These robots enable cost-effective inventory tracking by supporting inventory identification in put-away locations, near real-time analysis and visualisation of product storage. 
Recent reports highlight the significant social and economic benefits of using mobile robotics in industrial environments and the public domain~\cite{WRS,LRF}.

Despite the potential robotics-induced benefits,  more often than not, the robotic systems are underpinned by fragile designs leading to unavoidable obsolescence issues~\cite{jennings2016forecasting,gerasimou2017technical}. 
The reasons for this fragility are primarily \textit{technological}, resulting in an architectural gap between system components~\cite{alelyani2019literature}. 
The continuous development of new hardware and software components that provide capability and performance improvements reduces the long-term support of legacy components.
Although the generic architecture of modern robotic systems enables hardware components to be put together in a `plug and play' manner, hence, motivating system upgrade with more technologically advanced components, the significant cost incurred is an inhibiting factor~\cite{Sandborn2008:HPE}.
Other reasons contributing to fragile designs include changes in functional or non-functional requirements (such as new timing requirements) leading to incompatibility between system components.

Selecting an appropriate modernisation strategy is undoubtedly critical for reducing the risk of technology stagnation and modernising the robotic system successfully~\cite{JSPP886,IEC,rajagopal2015impact}. 
Pursuing a complete re-development strategy, albeit useful since it drives re-designing, refactoring and customisation of the legacy robotic system, incurs unrealistic costs that are directly proportional to the size of the system~\cite{fleurey2007model}. 
Similarly, adopting a conventional code-based migration approach is not only time-consuming but also error-prone since it ignores rich models (e.g., hardware architecture diagrams) that are typically produced by the design team of an organisation during the early stages of system development. 
In contrast, employing a \textit{model-based strategy} that facilitates the migration of the robotic system to a new hardware platform using available architectural models offers several benefits, including better maintainability, reduced cost and reusability~\cite{raibulet2017model}.


In this paper, we introduce \approach, a model-driven software migration approach that supports migration of the software controlling a robotic system between hardware platforms. 
First, 
\approach\ uses static analysis on the robotic software of the source hardware platform to extract platform-specific and platform-agnostic constructs. 
The static analysis is driven by the C/C++ Development Tooling (CDT) model driver of the Epsilon platform~\cite{kolovos2010epsilon} specifically developed to support \approach.
Through transformations of models capturing the hardware architecture of components on the target platform (e.g., temperature sensors, servo motors), \approach\ determines key attributes for those components (e.g., sensor type, interfaces used to communicate with the target platform).
For each component, \approach\ produces a ranked list of candidate software libraries suitable for the target platform, by employing the TFIDF statistical measure for information retrieval~\cite{salton1986introduction} and combining extracted information for this component and a platform-specific repository enhanced with historical data. 
Finally, \approach\
uses code generation to 
refactor the software, making it suitable for the target platform by realising the adapter pattern~\cite{gof}, highlight platform-specific constructs that 
need manual edits 
and add software constructs to support the invocation of hardware components on the target platform. 
Engineers can complete the migration by inspecting the highlighted areas
and adding source code that enables the invocation of hardware components on the target platform.


We evaluate \approach\ by migrating two robotic systems deployed for a line following and an environmental monitoring mission from a Propeller Activity Board to an Arduino Uno microcontroller~\cite{Warren2011}.
Our experimental evaluation shows that \approach\ can support the modernisation of robotic systems by making suitable modifications to the software running on the source platform.
The manual adaptations needed to complete the migration and develop a fully-fledged robotic system for Arduino focus on inferring constructs (i.e., invocations to Arduino libraries and auxiliary code) that achieve the same functionality as the source system. 
We did not observe any perceptible performance difference when deploying the robotic system on the target platform.



The main contributions of our paper are as follows:
\squishlist
\item The \approach\ approach enabling the migration of robotics software between hardware platforms (Section~\ref{sec:approach});
\item The Epsilon Model Connectivity (EMC) CDT model driver that enables the management of C/C++ software using MDE technologies specifically developed to support \approach;
\item The open-source \approach\ prototype tool presented in~ Section~\ref{sec:implementation} and made available on our project webpage;
\item The evaluation of the \approach\ approach and prototype tool summarised in Section~\ref{sec:evaluation}.
\squishend


%% file: sections/s2-problem.tex

\section{Motivating Example}\label{sec:example}



\begin{figure}[t]
	\centering
	\includegraphics[width=\linewidth]{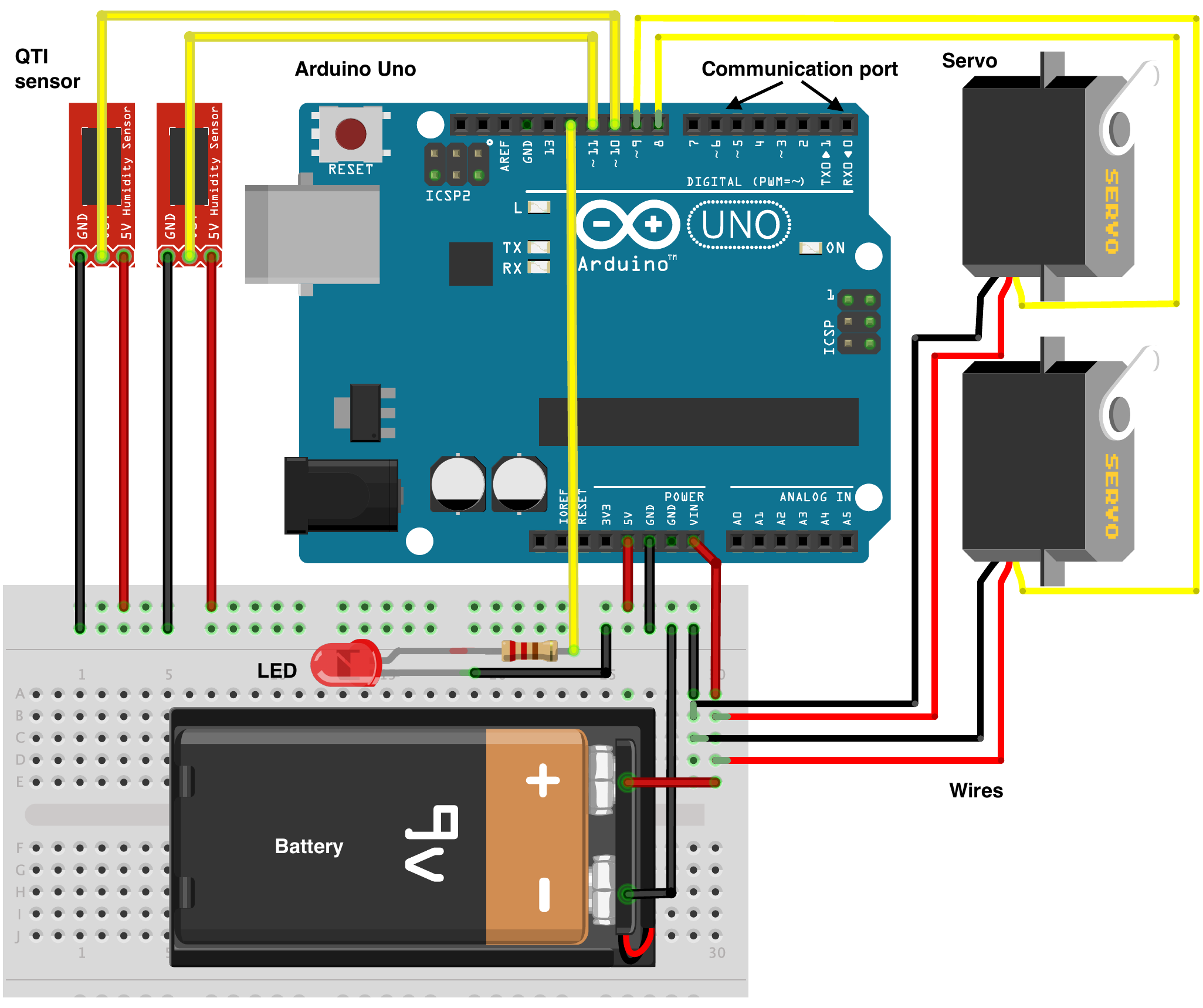}
	
	\vspace*{-2mm}
	\caption{Architecture of the line following robotic application in Fritzing.}
	\label{fig:example}	
	
	\vspace*{-4mm}
\end{figure}

\begin{figure}[t!]
	\begin{minipage}[t]{\linewidth}
		\begin{lstlisting}	[language=C++, label={lst:exampleCode},escapechar=|,  
		caption={Software excerpt of the line following robotic  application running on the Propeller Activity Bot.}\vspace*{1mm}]
		#include "simpletools.h"
		#include "abdrive.h"|\vspace*{1mm}|
		const int leftQTI  = 6;  //Left QTI port
		const int rightQTI = 7;  //Right QTI port		
		const int LED      = 8;  //LED port
		const int QTI_T    = 50; //QTI sensor threshold
		const int waitTime = 230;//QTI wait time |\vspace*{1mm}|
		long GetQTIState(int comPort){
			set_output(comPort, 0b1);
			set_direction(comPort, 0b1);
			waitcnt(waitTime);
			set_direction(comPort, 0b0);
			waitcnt(waitTime);
			long state = get_state(comPort);
			return state;
		}|\vspace*{1mm}|
		int main(){
			while (1){
				long sLeft  = GetQTIState(leftQTI);  
				long sRight = GetQTIState(rightQTI); |\vspace*{1mm}|							
				if (sLeft < QTI_T && sRight < QTI_T)
					drive_speed(24, 24); //continue straight
				else if(sLeft < QTI_T && sRight > QTI_T)
					drive_speed(24, 12); //turn left
				else if(sLeft > QTI_T && sRight > QTI_T){
					drive_speed(0, 0);  //stop
					high(LED);          //switch on LED
					pause(1000);         //wait 1s
					low(LED);           //switch off LED
					drive_speed(24, 24);//continue
				}
		}
		...\end{lstlisting}
		\vspace*{-4mm}
	\end{minipage}
\end{figure}

We will illustrate the \approach\ approach for the model-driven migration of robotics software using a robotic application deployed to navigate across the floor of a building by following a dark line. 
The robotic application has the target architecture shown in Figure~\ref{fig:example} developed using the open-source electronic computer-aided design (ECAD) tool Fritzing~\cite{knorig2009fritzing}.
Similar to other ECAD tools (e.g., KiCad, Eagle), Fritzing enables to design schematics of electronic-based prototypes by selecting hardware components (e.g., an Arduino Uno hardware platform, temperature sensors, servos, resistors) and wiring these components together through their communication ports to generate a closed circuit that performs a specific task. 
The generated schematics can be used to produce printed circuit boards for hardware-in-the-loop simulation or for manufacturing and commercialisation purposes.

The robotic application uses an Arduino Uno as its target hardware platform for controlling the execution of the application, moves using two servos and maintains its position on the dark line using two QTI  sensors. 
These sensors can identify light and dark surfaces (e.g., a black line drawn on a white floor) through using an infrared light-emitting diode (LED) and an infrared phototransistor internally.
The light emitted by the LED bounces off a surface and is detected by the phototransistor. 
Surfaces with darker colour absorb the emitted light, and thus the phototransistor gives lower values.
Once a checkpoint is discovered (i.e., both QTI sensors detect a dark mark), the robot emits light through the red LED. 
The resistor between the LED and Arduino is a passive component that protects the LED by limiting the flow of electrical current when a voltage is applied across it.
Finally, a 9V battery provides power to the robotic application. 

Listing~\ref{lst:exampleCode} shows an excerpt of the C software for this robotic application deployed on a Propeller Activity Bot. The initial segment of this software includes the platform-specific libraries through the include directives (lines 1-2) and the declaration of 
the communication ports for the robot components (lines 3-5). 
The $\code{GetQTIState}$ function emits infrared light (lines 9-10) and calculates the amount of light detected by a phototransistor when reflected off by a nearby surface by measuring the rate of charge transfer through the  phototransistor (lines 11-15). 
While running, the application executes the $\code{while}$ loop whose purpose is to determine the values of the QTI sensors (by invoking the $\code{GetQTIState}$ function) and, depending on whether these values are below or above an engineer-defined QTI threshold (line 6), to steer the robot accordingly (e.g., straight, left, right) by invoking the $\code{drive\_speed}$ function of the servos (lines 21-31).  
When the robot encounters a checkpoint, it stops and switches on the red LED for one second, before switching off the LED and continuing its journey.

Although this robotic application might seemingly look simple,  the manual migration of the software from the Propeller Activity Board to a more powerful hardware platform (e.g., an Arduino Uno) is a challenging activity. In particular, this activity involves analysing the original software to determine which code fragments must be migrated and which must be deleted, identifying platform-specific libraries and modifying the communication ports to match the new platform. In the following section, we present the \approach\ approach and illustrate its ability to automate important steps of the migration activity.

%% file: sections/s3-methodology.tex


\vspace*{-2mm}
\section{Approach}
\label{sec:approach}
\noindent

\begin{figure}[t]
	\centering
	\includegraphics[width=0.98\linewidth]{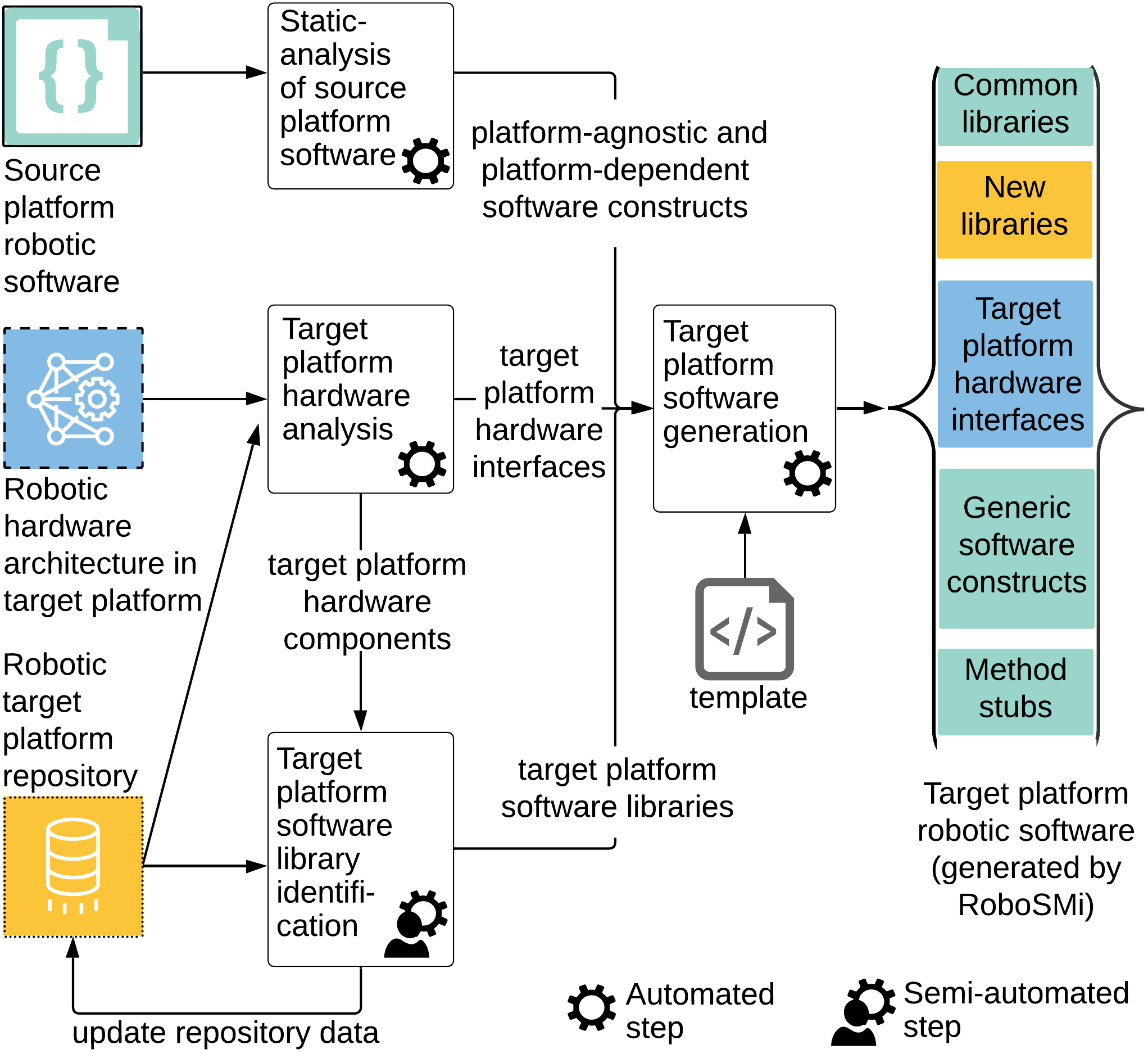}

	\vspace*{-2mm}
	\caption{\approach\ workflow.}
	\label{fig:approach}	

	\vspace*{-4mm}
\end{figure}

In this section, we describe our \approach\ model-driven approach for the systematic migration of software between robotic platforms. 
\approach, whose high-level workflow is shown in Figure~\ref{fig:approach},
uses the following artefacts: (i) the software deployed on the source robotic platform that will undergo the migration process (green solid box);  (ii) a model specifying the hardware architecture for the target robotic platform (blue dashed box); and (iii) a  platform-specific repository that holds key information about the target platform (yellow dotted box). We assume that no model of the source platform hardware is available.
This is a common issue for most modernisation activities as these legacy hardware specifications are often incompatible with currently used tools, might require tacit knowledge or have been misplaced.

\approach\ starts with analysing the software deployed on the source robotic platform to determine generic (i.e., platform-agnostic) and platform-dependent software constructs. 
These constructs form the input for the model-to-text transformation of software for the target platform. 
Next, through examining a hardware specification of the target robotic platform, \approach\ identifies information about the target hardware platform that is essential for the transformation. 
This set of information includes the types and characteristics of hardware components and interfaces used for communicating with the target platform. 
The next step of the approach involves combining the extracted information with information from the platform-specific repository to recommend suitable software libraries whose selection will enable the modernised code to access those hardware components. Finally, using all extracted key data from the previous three steps, the software for the target platform is generated through model-to-text transformation and realisation of the adapter design pattern~\cite{gof}. 
Although the generated code is fully compilable (i.e., an executable can be generated), software engineers should inspect it and complete the empty placeholders with suitable code (cf. Section~\ref{ssec:approachStep4}). 
In the following sections, we describe the general principles underpinning each step of \approach\ and present its realisation for the tool-supported \approach\ instance.

\subsection{Source Platform Software Static-Analysis}\label{ssec:approachStep1}
\noindent\textbf{Approach.} 
During this automated step, \approach\ analyses the software of the source platform to establish the set of modifications needed to make the software suitable for the target platform.
To this end, the source software undergoes a series of static analysis  	  actions 
including 
scanning and preprocessing, 
semantic analysis, name resolution and binding, that enables to generate 
its Abstract Syntax Trees (ASTs).
Once parsing finishes, the generated 
ASTs are analysed using AST visitor implementations, to detect platform-agnostic and platform-dependent software constructs. 
Platform-agnostic constructs 
will remain unchanged during the software generation for the target platform.
For instance, include directives that enable accessing standard C/C++ library functions such as containers (e.g., queue, set) 
and input/output streams (e.g., stdio, fstream) are common across both platforms and will not be modified. 
In contrast, platform-dependent constructs (e.g., methods native to the source platform) require manual adaptations. 
Consequently, these constructs are marked and will be handled accordingly in the model-to-text transformation step (cf. Section~\ref{ssec:approachStep4}).

\vspace*{2mm}
\noindent\textbf{Realisation.}
The tool-supported \approach\ instance carries out the static analysis step using a combination of lightweight model querying and transformation operations. 
More specifically, \approach\ uses the software as a model and navigates through it on-demand~\cite{izquierdo2014extracting}. Thus, the entire analysis and extraction of software constructs is guided through a set of model management queries at the abstract syntax level defined in the Epsilon Object Language (EOL)~\cite{kolovos2006epsilon}. 
This alleviates the need to transform the source platform software into an EMF-compatible representation, a task that not only requires accurate model extraction tools and a complete metamodel of the analysed general-purpose programming language, but it is also time-consuming for large software systems~\cite{bruneliere2014modisco}.

The \approach\ analysis is underpinned by the EMC CDT model driver~\cite{cdtDriver}, a ``technology-specific driver'' 
that exposes the document object model (DOM) maintained by the Eclipse C/C++ Developer Tools (CDT)~\cite{eclipseCDT} as models. 
Accordingly, the EMC CDT driver provides access to the internal representations maintained by CDT in the form of models that are suitable for the model management languages of the Epsilon platform~\cite{kolovos2006epsilon}. 
Once a query is made (e.g., to identify platform-independent include directives), the EMC CDT driver employs a ReflectiveASTVisitor, a specialisation of the visitor design pattern~\cite{gof}, 
to parse the relevant software and perform the necessary binding resolution, to traverse the DOM abstract syntax tree and, finally, to return all software constructs meeting the constraint (e.g., include directives). A similar process is applied for queries that extract platform-dependent constructs. \approach\ caches the results in memory to reduce parsing and analysis times of similar queries in the future.

\begin{example}
	Consider Listing~\ref{lst:exampleCode} of the robotic application (Section~\ref{sec:example}). 
	The functions $\code{set\_output}$, $\code{set\_direction}$, $\code{waitcnt}$, $\code{high}$, $\code{low}$, $\code{get\_state}$, $\code{drive\_speed}$,  and $\code{pause}$ are platform-dependent and are identified by \approach\ as constructs for modernisation.
	The include directives (lines 1-2) are specific to the Propeller platform and are not marked for the model-to-text generation step (cf. Section~\ref{ssec:approachStep4}). The remaining constructs (e.g., variable declarations on lines 5-10, function declarations on lines 12, 25 and 30) are platform-independent and will not be modified during the modernisation. 
\end{example}

\subsection{Target Platform Hardware Analysis}\label{ssec:approachStep2}
\noindent\textbf{Approach.}
In this \approach\ step, a diagrammatic specification of the system architecture for the target hardware platform is used for the extraction of key information of the components comprising the target platform. 
This is an important step as each platform has its own architecture with specific communication interfaces (e.g., ports) for hardware components (e.g., robotic servos), power access points, and software libraries for digital and analog input/output. 
Hence, the migration of the robotic application requires to correctly determine the type of used hardware components (i.e., models of sensors and actuators) and to establish the communication interfaces through which the software can interact with the components on the target platform. 
Since the objective is the generation of a functionally-equivalent application for the target platform, identifying the hardware components must be enhanced with information of communication interfaces with the target platform. 
The manual execution of these tasks is challenging and error-prone, especially 
when different teams undertake the generation of the diagrammatic specification and software development.
This is a common scenario in large organisations that causes the software team to spend a considerable amount of time to study the diagrammatic specification before proceeding with the migration. 
\approach\ automates this step by transforming the diagrammatic specification of the target platform configuration into an EMF compatible model that is suitable for model management operations. Given the corresponding EMF model, \approach\ executes model queries to extract both the precise model of employed hardware components (used in the next \approach\ step for software library recommendation) and the communication interfaces from the target hardware platform.

\begin{figure}[t]
	\centering
	\includegraphics[width=0.9\linewidth]{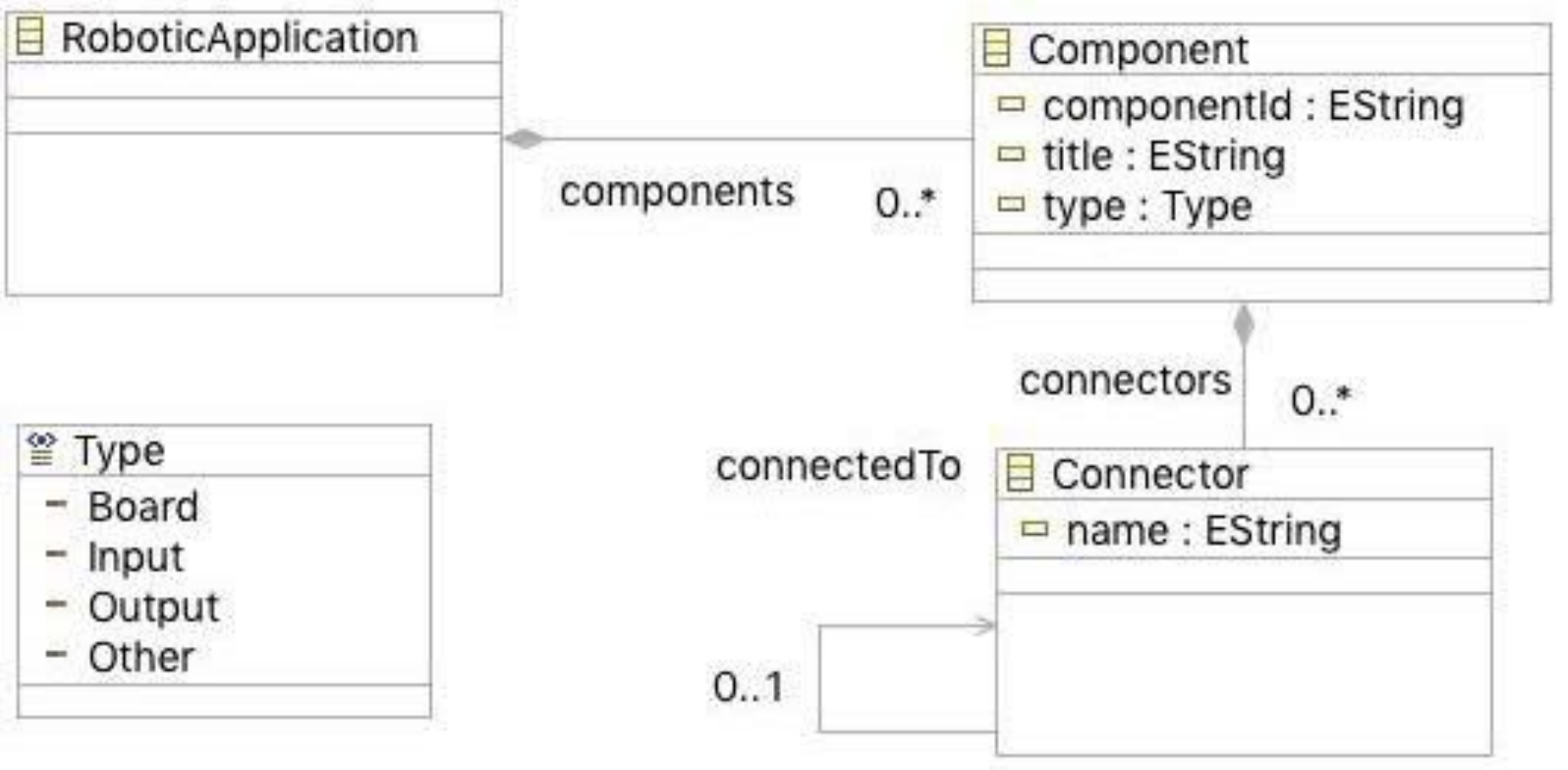}
	
	\vspace*{-2mm}
	\caption{Metamodel of a robotic application}
	\label{fig:metamodel}	
	
	\vspace*{-4mm}
\end{figure} 

\begin{figure}[t]
	\centering
	\includegraphics[width=\linewidth]{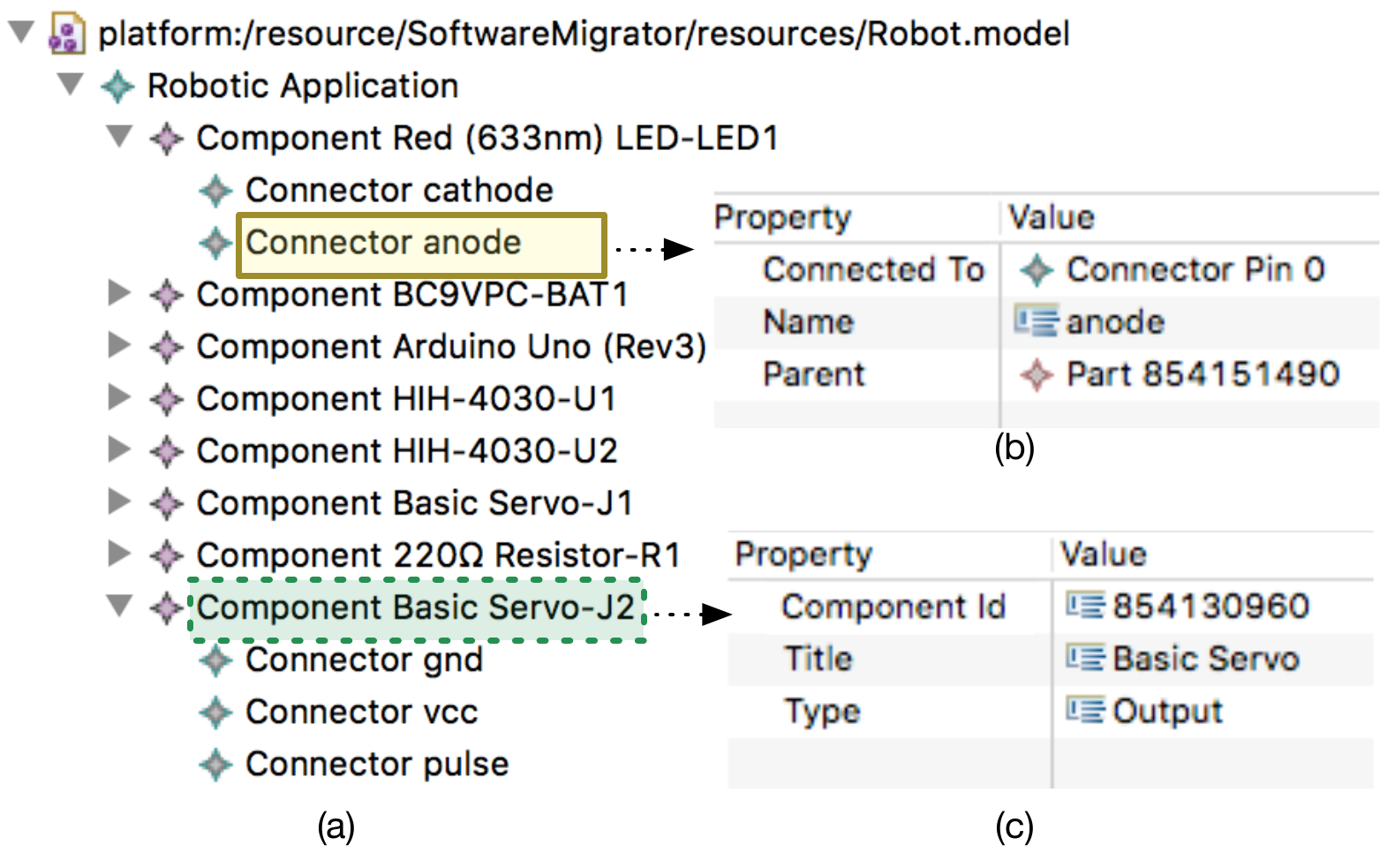}
	
	\vspace*{-4mm}
	\caption{Model instance (a) of the line following robotic application from Section~\ref{sec:example} corresponding to the metamodel from Figure~\ref{fig:metamodel} with information about the LED anode connector (b)  and the servo component (c).}
	\label{fig:qtimodel}
	
	\vspace*{-6mm}	
\end{figure}

\vspace*{2mm}
\noindent\textbf{Realisation.}
The tool-supported \approach\ instance uses as input a Fritzing specification~\cite{knorig2009fritzing} of the robotic application for the target hardware platform. Figure~\ref{fig:example} shows the Fritzing specification of the line following robotic application introduced in Section~\ref{sec:example}.
Given a Fritzing project as an XML file, \approach\ carries out a text-to-model transformation to 
instantiate the metamodel shown in Figure~\ref{fig:metamodel}.
This robot-specific metamodel represents the configuration of a robotic application as a set of components. Each component is assigned a specific type (e.g., Board, Input, Output) and has a set of connectors through which it can connect to (i.e., communicate) other components on the platform. 
The type of each component is extracted from a platform-specific repository which records whether the component provides input (e.g., a temperature sensor), output (e.g., a servo) or auxiliary (e.g., a resistor) functionality. 
The main component is the hardware platform that runs the robotic application and is of type Board. The connectors of this component include all the digital and analog interfaces through which the platform can communicate with other components. 
Extracting this information by directly analysing the XML file (e.g., through the XML driver for Epsilon~\cite{kolovos2010epsilon}), albeit possible, it introduces additional complexity, requires redundant analyses of the XML file,
and would make the combination of information from the platform-specific repository convoluted. 
Figure~\ref{fig:qtimodel} shows the EMF model for the line following robotic application (Figure~\ref{fig:example}) along with extracted information for the hardware components and connectors.

Having generated the EMF model instance of the Fritzing specification, \approach\ analyses thoroughly the model to determine a mapping set that signifies how connectors from the hardware platform are connected with components that perform input and output functionality (i.e., sensors and actuators).
Setting the connectors correctly is important both for having a valid hardware specification and for enabling the software to access and control the hardware components. Incorrect connector specification could damage the hardware components or the entire hardware platform.

\approach\ executes Algorithm~\ref{alg:connectorMapping} to generate the mappings. 
Given the connectors on the hardware platform by the $\textsc{GetConnectors}$ function (line~\ref{al:l3}), \approach\ selects each connector $c$ in sequence and analyses the transitive association with other connectors until it reaches a connector whose parent is an (active) input or output component (lines~\ref{al:l10}--\ref{al:l11}). 
To establish this transitive association, \approach\ applies a depth first search strategy as only connectors leading to input or output components are really accessible through the software. 
In contrast, passive components (e.g., resistors) should be traversed to determine the input/output components with which they are connected. 
For instance, resistors are necessary for limiting the flow of electrical current in an electronic circuit or for providing a specific voltage to another component (e.g., temperature sensors, servos).
Despite their importance, 
the value of these components in the software side is on identifying the input/output components they ``protect'' through analysing their connectors. 
Thus, when these passive components are encountered during model traversal, \approach\ adds their connectors to the set of connectors for analysis (lines~\ref{al:l13}--~\ref{al:l15}). 
Algorithm~\ref{alg:connectorMapping} can also establish the mappings with 	components requiring multiple connections to the hardware platform (e.g., a liquid crystal display requires six connections).
Once the algorithm completes, it returns the $\mathsf{MAPPINGS}$ data structure that denotes how connectors from the hardware platform are connected to connectors from input/output components (line 16). 
\\\noindent\textbf{Complexity analysis.} We can define the system architecture as a graph with $N$ nodes and $E$ edges. The nodes $N^C \subseteq N$ belong to the hardware platform $P$. For each $c \in N^C$, Algorithm~\ref{alg:connectorMapping} executes a depth-first search until finding an edge whose ending node is an input or output component. Consequently, the worst-case performance for graph traversal without repetition  is $N^C \times O(|N| + |E|)$.  

\begin{example}\label{ex:mappings}
	Consider the Fritzing specification of the line following robotic application from Figure~\ref{fig:example}. The execution of this \approach\ step generates the model in Figure~\ref{fig:qtimodel}. The text next to a component corresponds to its Fritzing-specific unique identifier. Algorithm~\ref{alg:connectorMapping} produces the $\mathsf{MAPPINGS}$ structure \{BasicServo2WriteP1$\rightarrow$8, BasicServo1WriteP1$\rightarrow$9, HIH40302ReadP1$\rightarrow$10, HIH40301ReadP1 $\rightarrow$11, Red633nm-LED1WriteP1$\rightarrow$12\} that maps a unique identifier for each connector of every component to the hardware platform connector it is connected. 
	For instance, \{Red633nm-LED1WriteP1$\rightarrow$12\} denotes that the anode connector of the red LED (Figure.~\ref{fig:qtimodel}a) is connected to connector \#12 of Arduino Uno. This mapping has been established through traversing the resistor's connectors between the LED and Arduino. 	
\end{example}

\begin{figure}[t]
		\vspace*{-3mm}
	
\begin{minipage}[t!]{\linewidth}
\begin{algorithm}[H]
	\caption{Platform component identification}\label{alg:connectorMapping}
	\begin{small}
		\begin{algorithmic}[1]
			\Function{AnalyseHardwarePlatform}{P}
			\State $\mathsf{MAPPINGS} \gets \emptyset$
			\ForAll{$c \in \textsc{GetConnectors}(P)$}\label{al:l3}
			\State S $\gets \{c\}$
			\State V $\gets \emptyset$
			\While {$\neg (S=\emptyset$)}
			\State d $\gets \textsc{SelectFirst}(S)$
			\If {$d \notin V$}
			\State $V = V \cup \{d\}$
			\If {$\textsc{Type}(\textsc{Parent}(d)) \in \{Input,Output\}}$\label{al:l10}
			\State $\mathsf{MAPPINGS} \gets \textsc{Append}(c, d)$\label{al:l11}
			\Else
			\State $ S \gets  S \cup \{\textsc{ConnectedTo}(d)\}$
			\IIf {$\textsc{Type}(\textsc{Parent}(d))\notin \{Board\}$}\label{al:l13}
			\State $\hspace*{5mm}S \gets S \cup \textsc{GetConnectors}({\textsc{Parent}(d)})$\label{al:l15}
			\EndIIf
			\EndIf
			\EndIf
			\EndWhile
			\EndFor \label{a1:l15}
			\State  \textbf{return} $\mathsf{MAPPINGS}$
			\EndFunction
		\end{algorithmic}
	\end{small}

\end{algorithm}
\end{minipage}
	\vspace*{-6mm}
\end{figure}

\subsection{Target Platform Software Library Identification}\label{ssec:approachStep3}
\noindent\textbf{Approach.} 
In this \approach\ step, candidate software libraries for every hardware component of the target platform are identified and recommended to engineers. 
Since hardware platforms are typically packaged with several
software libraries\footnote{e.g., https://www.arduino.cc/enReference/Libraries} 
that enable interacting with components deployed on the platform, 
selecting the most suitable library for each component by manually inspecting those libraries is a non-trivial task. 

\begin{figure}[t]
	\centering
	\includegraphics[width=0.85\linewidth]{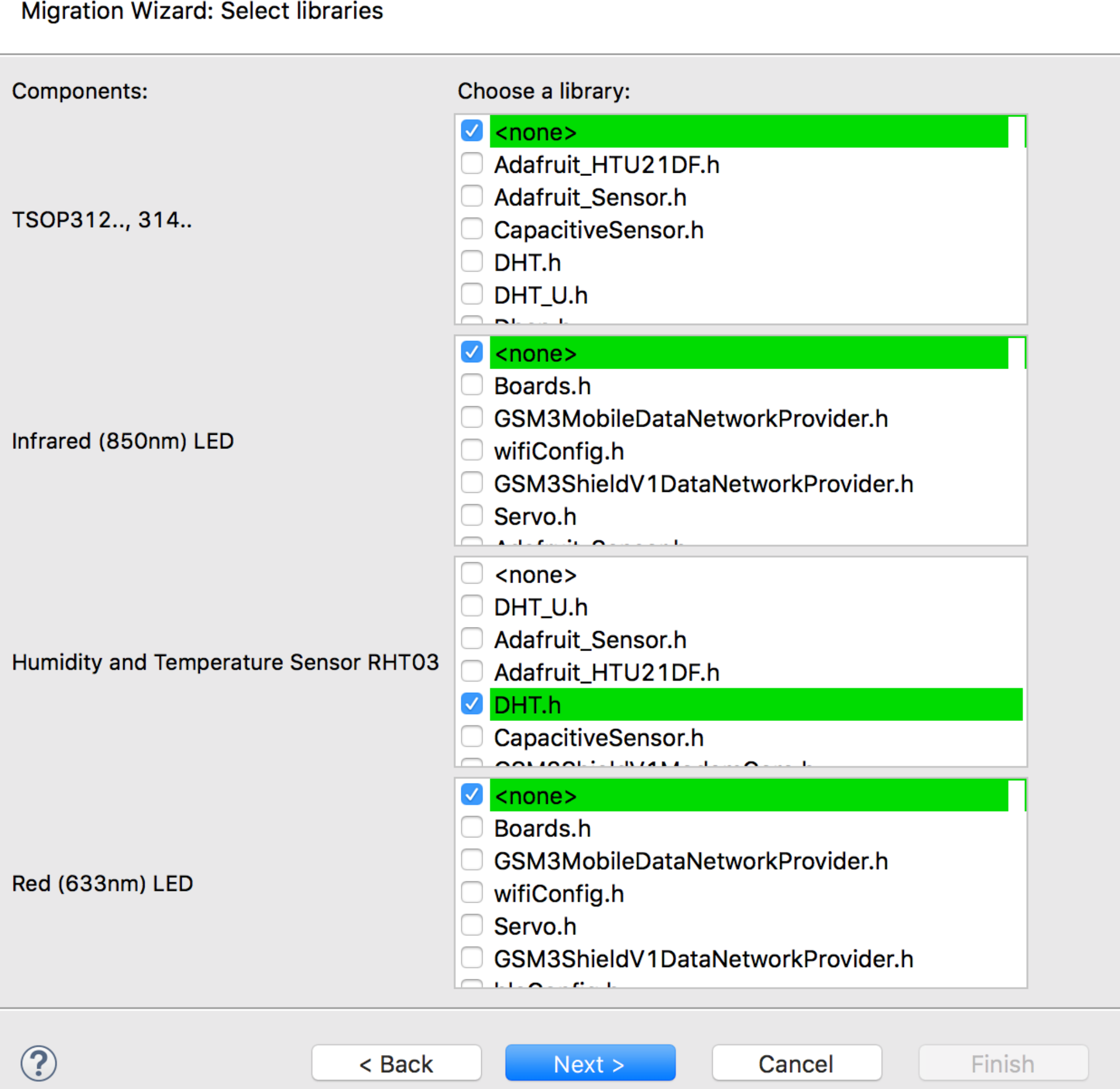}
	\vspace*{-2mm}
	\caption{Ranked software libraries per component for the line following robotic application from Section~\ref{sec:example}.}
	\label{fig:lib_selection}	
	
	\vspace*{-6mm}
\end{figure}

\approach\ facilitates software library selection by recommending a set of candidate software libraries for the target platform for each hardware component by exploiting the information available about the software libraries and the components. More specifically,  
\approach\ combines key information for each component identified through model analysis of the Fritzing specification (Section~\ref{ssec:approachStep2}) with information about software libraries for the target platform and historical data of previously used libraries for the component, both available in the target platform repository. 
After the most likely software libraries for each component have been identified, they are presented to engineers through an intuitive user interface~(Figure~\ref{fig:lib_selection}) for selecting the library to realise the functionality of the component. 
Once the selection task is completed, information about the selected libraries is forwarded to the final \approach\ step for the generation of the target platform software.

\vspace*{2mm}
\noindent\textbf{Realisation.}
The tool-supported \approach\ instance produces a ranked list of candidate software libraries for each hardware component using a two-phase method. 
During the first phase, \approach\ transforms the ranking into an information retrieval task that aims at establishing the relevance score of a library based on a component's name and a collection of header files corresponding to software libraries available for the target platform. 
A header file holds software constructs of a library (e.g., function and variable declarations) and represents its publicly available interface via which third-party client software (e.g., the software for the target platform) can invoke the functionality of the library. 
Given a hardware component, \approach\ 
establishes for each library its 
term frequency-inverse document frequency (\emph{TFIDF}) score~\cite{salton1986introduction} given by
\begin{equation}\label{eq:tdidf}
\textsc{tdidf}_d=
\sum_{t\in T} \big(1 + log(tf_{f,d})\big) \times log\bigg(\frac{N}{df_t}\bigg)
\end{equation}

\noindent
where
\hspace*{2mm}$T:$ the set of terms\\
\hspace*{11mm}$df_t:$ the number of documents term $t$ occurs in\\
\hspace*{11mm}$N:$ the number of documents in the collection\\
\hspace*{8mm}$tf_{t,d}:$ occurrences of term $t$ in document $d$\\

TFIDF weighs a term’s frequency (TF) and its inverse document frequency (IDF) with each term having its respective TF and IDF score.
In \approach, a document corresponds to a header file and a term corresponds to a part in the name of a component. The total ranking for a library is calculated by summing the TFIDF score for each term in the name.
TFIDF takes into account both the frequency of occurrence of a term in the library as well as its ``informativeness''. As such, a term is more informative if it appears in fewer documents of the collection; ``informativeness'' effectively corresponds to the $log\Big(\frac{N}{df_t}\Big)$ term in~(\ref{eq:tdidf}). 
For example, if we consider the QTI sensor in the line following robotic application, the term \textit{sensor} is less informative than the term \textit{QTI} as many components have the term \textit{sensor} in their name and so will many header files. 
If ``informativeness'' is not taken into account, any component that contains the term \textit{sensor} in its name is likely to rank higher than the QTI sensor component, since the term \textit{sensor} will have the most occurrences in the text, regardless of the type of sensor. 
Since \textit{QTI} is a less common term, using the TFIDF score will instead rank header files containing the less frequently occurring term \textit{QTI} higher.

The accuracy of rankings depends on the most suitable libraries for a component making some reference or including similar terms to the component’s name. 
Given the information provided in the Fritzing specification (cf. Section~\ref{ssec:approachStep2}), the name of a component is the only information that could be used for determining the TFIDF score. 
Exploring other statistical measure techniques and improving the accuracy of TDIDF  is part of our future work (cf. Section~\ref{sec:conclusion}).

To improve further the quality of library recommendations, \approach\ in its second phase executes a set of queries to a platform-specific database of historical configuration data 
to establish whether the component has been used previously in another migration activity. If this holds, \approach\ extracts the chosen libraries for this component along with a counter indicating how many times each library has been chosen. The higher the number, the more likely this library will be selected for this migration activity.

By using a utility function that combines the TDIDF scores and the information from the historical configuration data, \approach\ generates a ranked list of candidate libraries for each hardware component. 
Engineers can review this list and select the libraries matching the system's components (Figure~\ref{fig:lib_selection}).
Once a selection has been made, the configuration data is updated accordingly. 
Since the correct library is expected to be chosen most frequently, over time this historical configuration data will improve the accuracy of library suggestions.

\vspace*{-4mm}
\subsection{Target Platform Software Generation}\label{ssec:approachStep4}
\noindent\textbf{Approach.}
The last \approach\ step  involves the automatic generation of software for the target platform by instantiating a software template using information extracted from the previous three steps (see Figure~\ref{fig:approach}). 
In particular, for each source file provided as input from the source platform, \approach\ generates the corresponding source file for the target platform in which platform-agnostic software constructs are retained (Section~\ref{ssec:approachStep1}), information about the ports for interfacing with the hardware components is added (Section~\ref{ssec:approachStep2})
as well as include directives for the selected software libraries (Section~\ref{ssec:approachStep3}). 
For platform-specific constructs, extracted from analysing the source platform software (Section~\ref{ssec:approachStep1}), \approach\ implements the adapter design pattern~\cite{gof} and generates placeholders using the signatures of these constructs enhanced with suitable TODO directives. 
Using this pattern reduces the changes to the source software while delegating the modifications to start exercising the functionality on the target platform to the produced placeholders~\cite{tonelli2010swing,bartolomei2009study}.
This pattern has also been applied for the integration of legacy systems using MDE~\cite{clavreul2010integrating}.
Having a fully working software for the target platform requires to populate the empty placeholders with suitable code (e.g., invocations to methods of selected libraries).
The TODO directives, summarised into a task list, provide software engineers with a clear view of parts of the generated software that require manual inspection, refactoring or completion to finalise the migration to the target platform.

\vspace*{1mm}
\noindent\textbf{Realisation.} 
The tool-supported \approach\ instance generates the software for the target platform by executing a model-to-text transformation in the Epsilon Generation (EGL) Language~\cite{rose2008epsilon}. Listing~\ref{lst:eglTemplate} shows an excerpt of the transformation that generates the include directives for the platform-agnostic libraries (lines 1-3), the constructs for communicating with the hardware components (lines 4-7) and the implementation of the adapter pattern for the platform-specific methods found (lines 8-12).

\begin{figure}[t]
	\begin{minipage}[t]{\linewidth}
		\begin{lstlisting}[mathescape, language=EGL, label={lst:eglTemplate},  
		caption={\hspace*{1mm}EGL template except for generating the target platform software.}\vspace*{1mm},escapechar=|]
		[*Add platform-agnostic libs*]
		[$\%$ for (i in %getPlatformAgnosticLibraries()%){$\%$]
		#include ``[$\%$=i$\%$]''[$\%$}$\%$]|\vspace*{1mm}|
		[*Init port interfaces*]
		[$\%$ var ports = %generatePortInterfaces()%;$\%$]  
		[$\%$ for (p in ports.keySet()) {$\%$]
		const int [$\%$=p$\%$] = [$\%$=ports.get(p)$\%$]; [$\%$}$\%$]|\vspace*{1mm}|
		[*Add platform-dependent constructs*]
		[$\%$ for (d in %getPlatformDependConstructs()%){$\%$]
		[$\%$= d.getDeclarator().getRawSignature() $\%$]
		{\n//TODO: complete method\n}
		[$\%$}$\%$]\end{lstlisting}
		
		\vspace*{-6mm}
	\end{minipage}
\end{figure}

\begin{figure}[t]
	\begin{minipage}[t]{\linewidth}
		\begin{lstlisting}	[language=C++, label={lst:migratedCode},  
		caption={\hspace*{1mm}Generated software excerpt of the line following robotic application (the colouring scheme corresponds to the semantics in Figure~\ref{fig:approach})\vspace*{1mm}},escapechar=|]
		|\makebox[0pt][l]{\color{greenCode!40}\rule[-0.6ex]{29em}{3mm}}|#include <stdio.h>
		|\makebox[0pt][l]{\color{blueCode!40}\rule[-0.6ex]{29em}{3mm}}|#include "Arduino.h"
		|\makebox[0pt][l]{\color{blueCode!40}\rule[-0.6ex]{29em}{3mm}}|#include <Servo.h>
		|\makebox[0pt][l]{\color{yellowCode!30}\rule[-0.6ex]{29em}{3mm}}|const int QTI_T    = 100; //QTI sensor threshold
		|\makebox[0pt][l]{\color{yellowCode!30}\rule[-0.6ex]{29em}{3mm}}|const int leftQTI     = 10; //Left QTI port
		|\makebox[0pt][l]{\color{yellowCode!30}\rule[-0.6ex]{29em}{3mm}}|const int rightQTI    = 11; //Right QTI port		
		|\makebox[0pt][l]{\color{yellowCode!30}\rule[-0.6ex]{29em}{3mm}}|const int leftSrvPrt  = 8;  //Left servo port
		|\makebox[0pt][l]{\color{yellowCode!30}\rule[-0.6ex]{29em}{3mm}}|const int rightSrvPrt = 9;  //Right servo port
		|\makebox[0pt][l]{\color{yellowCode!30}\rule[-0.6ex]{29em}{3mm}}|const int LED         = 12; //LED port
		|\makebox[0pt][l]{\color{greenCode!40}\rule[-0.6ex]{29em}{3mm}}|long GetQTIState(int comPort){
		|\makebox[0pt][l]{\color{greenCode!40}\rule[-0.6ex]{29em}{3mm}}|	...
		|\makebox[0pt][l]{\color{greenCode!40}\rule[-0.6ex]{29em}{3mm}}|}
		|\makebox[0pt][l]{\color{white!10}\rule[-0.6ex]{29em}{3mm}}|void setup() {//Arduino specific|\hspace*{-1mm}| (startup)
		|\makebox[0pt][l]{\color{white!10}\rule[-0.6ex]{29em}{3mm}}|	...
		|\makebox[0pt][l]{\color{white!10}\rule[-0.6ex]{29em}{3mm}}|}
		|\makebox[0pt][l]{\color{white!10}\rule[-0.6ex]{29em}{3mm}}|void loop() {//Arduino specific|\hspace*{-1mm}| (running forever)
			|\makebox[0pt][l]{\color{greenCode!40}\rule[-0.6ex]{27.8em}{3mm}}|int sLeft  = GetQTIState(leftQTI);  
			|\makebox[0pt][l]{\color{greenCode!40}\rule[-0.6ex]{27.8em}{3mm}}|int sRight = GetQTIState(rightQTI);
			|\makebox[0pt][l]{\color{greenCode!40}\rule[-0.6ex]{27.8em}{3mm}}|if (sLeft < QTI_T && sRight < QTI_T)
			|\makebox[0pt][l]{\color{greenCode!40}\rule[-0.6ex]{27.8em}{3mm}}|	drive_speed(24, 24); //continue straight
			...				
		|\makebox[0pt][l]{\color{white!10}\rule[-0.6ex]{29em}{3mm}}|}
		|\makebox[0pt][l]{\color{greenCode!40}\rule[-0.6ex]{28.4em}{3mm}}|void drive_speed(int left, int right){
			|\makebox[0pt][l]{\color{greenCode!40}\rule[-0.6ex]{27.8em}{3mm}}|//TODO: complete method
		|\makebox[0pt][l]{\color{greenCode!40}\rule[-0.6ex]{28.4em}{3mm}}|}		
		...\end{lstlisting}
		\vspace*{-8mm}
	\end{minipage}
\end{figure}

\vspace*{1mm}
\begin{example}
	Consider again the Propeller Activity Board software shown in Listing~\ref{lst:exampleCode}. The model-to-text transformation generates the software excerpt shown in Listing~\ref{lst:migratedCode}. 
	The colouring scheme matches the colours from the \approach\ workflow (Figure~\ref{fig:approach}) and signifies the output of the \approach\ step from which the corresponding code has been derived. For instance, the include directives (lines 2-3; shown in blue) correspond to the libraries selected in Figure~\ref{fig:lib_selection} and the \code{drive\_speed} method definition (lines 28-30) represents the platform-specific construct derived from source platform software analysis (Section~\ref{ssec:approachStep1}).  
	The \code{setup} and \code{loop} methods are required by the Arduino Uno platform and are included in the EGL transformation (omitted from Listing~\ref{lst:eglTemplate} due to space constraints).
\end{example}

%% file: sections/s4-implementation.tex

\section{\approach\ Prototype Eclipse Plugin}\label{sec:implementation}
To automate the software migration process, we implemented a prototype tool as an Eclipse plugin with the architecture shown in Figure~\ref{fig:approach}. 
Our \approach\ tool uses the Epsilon framework~\cite{kolovos2010epsilon} in all its components 
and for all model management tasks including model querying and transformations.
Finally, the static analysis component uses the Eclipse CDT Project~\cite{eclipseCDT} and Epsilon CDT driver~\cite{cdtDriver} for parsing the abstract syntax tree of the source platform software and using C/C++ models with Epsilon. 
The open-source \approach\ code, the full experimental results summarised in the following section, additional information about \approach\ and the case
studies used for its evaluation are available as open-source at
\url{https://github.com/gerasimou/RoboSMi}.

%% file: sections/s5-evaluation.tex
\noindent

\section{Evaluation}
\label{sec:evaluation}
\noindent
\subsection{Research Questions}

\vspace*{1mm}\noindent
\textbf{RQ1 (Validation): }
\textbf{Can \approach\ support migration between hardware platforms? }
We used this research question to establish if \approach\ can generate the required artefacts and support the modernisation of robotic systems. 

\vspace*{1mm}\noindent
\textbf{RQ2 (Automation Level):}
\textbf{What is the level of automation supported by \approach\ for this migration activity?}
Since \approach\ aims to reduce the effort required for software migration in robotic systems, we examined the manual code that remains to be written to complete the migration task.

\vspace*{1mm}\noindent	
\textbf{RQ3 (Performance):}
\textbf{What is the impact of the adapter pattern in software migration?}
We analyse the extent to which the placeholders generated due to the adapter pattern affects the completion of the migration to the target platform.

\vspace*{1mm}\noindent	
\textbf{RQ4 (Generality):}
\textbf{What is the generality level of \approach?}
We elaborate on the extent to which the \approach\ approach can support  software migration tasks in similar application domains.

\subsection{Experimental Setup}\label{ssec:experiments}

\begin{figure}[t]
	\centering
	\includegraphics[width=0.6\linewidth]{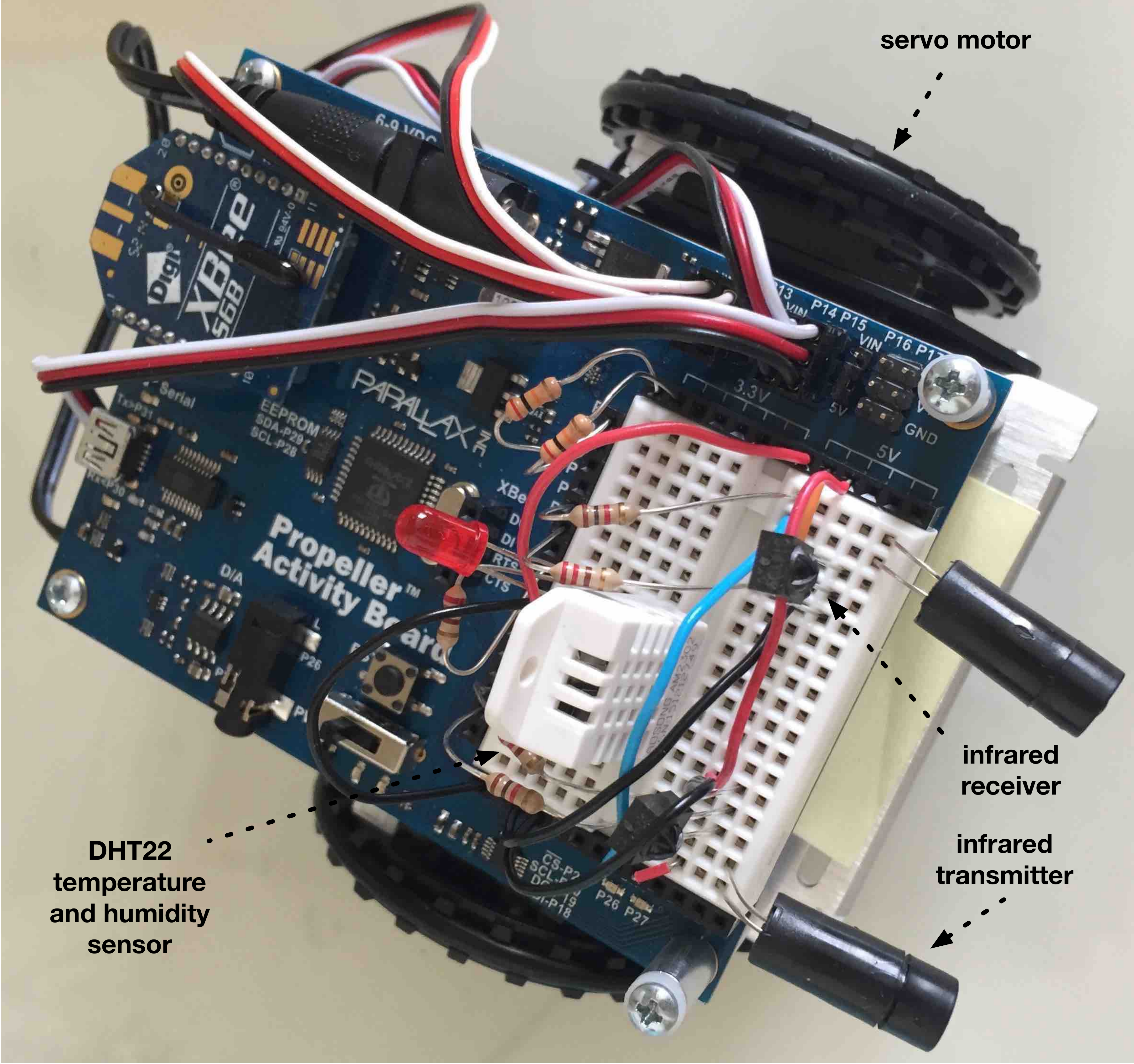}
	
	\vspace*{-2mm}
	\caption{Environmental monitoring robotic application running on the Propeller Activity board (https://www.parallax.com/product/32912).}
	\label{fig:monirotingRobot}	
	
	\vspace*{-4mm}
\end{figure}

We evaluate  the end-to-end \approach\ behaviour 
using two robotic applications similar to those available on 
open-source robotic repositories (e.g., \url{https://create.arduino.cc/projecthub}). 
The first is the line following application from Section~\ref{sec:example}. The second involves a robot deployed for an environmental monitoring mission. 
In particular, the robot, shown in Figure~\ref{fig:monirotingRobot}, can navigate its surroundings using its servo motors until a system of infrared transmitters and receivers detects objects blocking its path. 
When an obstacle is detected, the robot stops, rotates and continues moving in another direction.
Also, the robot takes periodic humidity and temperature readings of its operating environment using a DHT22 sensor. When any of these environmental attributes is above a predefined threshold, an LED is triggered to alert any interested stakeholders. 
The Fritzing specification is available on the project webpage.

Both applications are originally deployed on robots available in our lab using a Propeller Activity board and are migrated to functionally-equivalent robots using an Arduino Uno. 
As such, the source platform software contains a mixture of Propeller library functions, as well as global variables, user-defined functions, and include statements for non-Propeller libraries.
The hardware configuration in both platforms has the same hardware components but   
with  some variations in the communication ports with which components connect to platforms. This is due to platform-specific constraints, e.g., port \#12 on the Propeller is reserved for the left servo encoder whereas this port can support any component on an Arduino.

\subsection{Results and Discussion}
\noindent\textbf{RQ1 (Validation).}
We carried out the experiments described in the previous section and confirmed that \approach\ produced the artifacts for all its steps as presented in Sections~\ref{ssec:approachStep1}--\ref{ssec:approachStep4}. 
More specifically, the source platform software analysis step partitioned the software for both applications into Propeller-specific commands (e.g., include directives, method invocations to Propeller software libraries) and platform-agnostic commands (e.g., global variable declarations, locally defined and used variables and  methods).

By inspecting the Fritzing specification for each application, \approach\ generated the EMF model instances with characteristics of the hardware components (cf. Figure~\ref{fig:qtimodel}) and established the mappings set of connections between Arduino Uno (target platform) and the components. 
An interesting challenge we dealt with in this step was the extraction of a component's type (e.g., Input, Output) due to ambiguous and inconsistent information in the Fritzing specification. 
For instance, the distinguishing attribute in the Fritzing specification 
for the QTI sensor, DHT22 temperature and humidity sensor, infrared receiver, and servo motor had the values ``out", ``data-signal'', ``data'', and ``pulse'', respectively. Consequently, establishing the functionality type provided by a component using only this information is not trivial. We mitigated this challenge by extracting information from an Arduino-specific repository, incrementally developed during the course of the experiments.

Formulating the identification of mappings set in this second \approach\ step as a graph traversal problem and solving it using Algorithm~\ref{alg:connectorMapping} enables to address several challenges. 
First, by considering passive hardware components (e.g., resistors, potentiometers) as branch points (i.e., with at least one possible outgoing path) and applying a depth-first strategy, we establish the transitive association between ports on input/output components and the platform (cf. Example~\ref{ex:mappings}). 
Furthermore, the graph traversal problem formulation facilitates the identification of ports for components that require multiple ports on the platform. 
For example, an LCD needs six ports from the platform (four for data, and two for enabling and activating the component).
The depth-first strategy is also particularly useful when ports on the target platform are common among multiple components. 
This scenario can occur when the components are chained together and controlled by the same port. 
For the environmental monitoring application, for example, the port controlling the LED could be also used by a piezoelectric speaker that starts emitting sound upon exceeding the predefined threshold. 
Components are also chained together when employing a communication protocol other than the general-purpose input/output (GPIO), e.g., serial peripheral interface (SPI), inter-integrated circuit (I2C).
For instance, the I2C protocol uses a master-slave setup and requires only two ports to connect up to 1023 components (slaves) onto the same platform (master). 
Thus, \approach\ Algorithm~\ref{alg:connectorMapping} also automates the identification of ports for components using different communication protocols (GPIO, SPI, and I2C). 


The list of candidate software libraries produced in the third \approach\ step (cf. Section~\ref{ssec:approachStep3}) ranked the correct libraries among the top five (out of 247 in total) per application. 
Thus, the transformation of library identification into an information retrieval task and the calculation of TFIDF-based (\ref{eq:tdidf}) 
relevance scores provided practicable recommendations. 
For instance, the correct libraries \code{Servo.h} and \code{DHT.h} for the servo motor and the DHT22 sensor in the environmental monitoring application are ranked second and fourth, respectively. 
For DHT22, the more highly suggested libraries are all for humidity/temperature sensors, although not for this specific sensor. 

Since TDIDF combines information from the name of a component and the contents of the library's header, its relevance score presents interesting sensitivity attributes.
In particular, the name given to a component by Fritzing is particularly important. 
For example, the infrared receiver is named as ``TSOP312..'' which corresponds to a vendor-specific component and gives little indication on the functionality 
provided by the component. Consequently, the recommendations for this component are not useful. 
Another example concerns the servo motor for which the \code{Boards.h} library is ranked higher than \code{Servo.h}. 
This occurs because  \code{Boards.h} contains macro definitions (e.g., \code{MAX\_SERVOS}) for different types of boards (e.g. Arduino Mega, Uno). 
Hence the term ``servo'' appears many times, contributing to a higher TFIDF score. 
The use of historical configuration data partially addresses these issues.
If the correct library has been chosen in the past, it will be ranked higher in subsequent \approach\ use.

The instantiation of the adapter pattern to generate the Arduino Uno software in the final \approach\ step (cf. Section~\ref{ssec:approachStep4}) produces the expected output. 
To this end, suitable include directives are defined for 
platform-agnostic libraries and those selected in the previous step (cf. Section~\ref{ssec:approachStep3}), and placeholder methods with TODO directives are generated for any\hspace*{-0.1mm} invoked\hspace*{-0.1mm} Propeller-specific\hspace*{-0.1mm} library\hspace*{-0.1mm} method. Also,\hspace*{-0.3mm}
commands from the main method outside the \code{while(1)} loop are copied to the \code{setup} method,
commands within \code{while(1)} are copied to the \code{loop} method,
any other defined methods and global variables are copied to the generated software.

\vspace*{1mm}\noindent
\textbf{RQ2 (Automation Level).} 
To answer this research question, we completed the migration and developed fully-fledged robotic applications for the Arduino platform by manually
adapting the \approach\ outputs and populating the placeholders with suitable commands.
We confirmed the functional equivalence of the migrated applications empirically and extracted the following useful observations.

Differences in the architecture and the programming languages 
between the platforms require to handle the migration carefully. 
In contrast to Propeller that uses a lightweight C (due to limited memory - max memory is 32KB), Arduino supports full C/C++ commands. 
Consequently, software libraries that enable accessing components on Arduino (e.g., servo motor, DHT22 sensor) need additional declarations and supporting commands for object instantiation and initialisation, respectively. 
For instance, Arduino requires to define a servo variable globally (e.g., \code{Servo servoLeft()}) and perform extra tasks (in the \code{setup} method) to inform the servo object for the port that enables controlling the servo motor (e.g., \code{servoLeft.attach(leftSrvPrt)}). Using the servo motor on Propeller, however, needs no initialisation because the platform architecture expects that the left servo is attached to port \#12. 
Likewise, instantiating the DHT22 sensor object on Arduino requires both the port and DHT model (e.g., \code{DHT(DHTPort,\!\!\! DHTModel)}) as the DHT library is used for multiple models, while Propeller does not need a sensor object.

The level of difficulty for inferring commands (i.e., invocations to Arduino libraries and auxiliary code) that achieve the same functionality as the original method depends on the correspondence between the platform libraries~\cite{Robillard2013TSE,Dig2005ICSM}. 
When the correspondence is one-to-one, inferring the mappings is fairly simple.
For instance, the commands \code{high(port)} and \code{digitalWrite(port, HIGH)} switch on an LED on Propeller and Arduino, respectively. 
In most cases, however, complex one-to-many and many-to-one mappings were needed.
For example, the infrared transmitter on Propeller needs a single \code{freqout()} command to transmit the signal for a period of time, whereas on Arduino  two commands should be used (i.e., \code{tone()} and \code{delay()}). 
Similarly, receiving a humidity reading on Propeller requires to trigger a reading from the DHT22 sensor and then to retrieve the humidity value, in contrast to a single command needed on Arduino. 

Mappings inference is significantly more challenging in cases involving semantic differences between platform libraries. 
This discrepancy occurs for the libraries controlling the servos on the examined platforms. Moving the robot on Propeller requires a single invocation of the \code{drive\_speed()} method (e.g, line 24 in Listing~\ref{lst:exampleCode}) that sets each servo to a certain speed in ticks per second.
In contrast, the corresponding \code{servoLeft.write()} method from the Arduino servo library 
enables to control each servo individually by defining the exact pulse width in microseconds. 
Therefore, additional effort was needed to find suitable transformations in these situations.

\vspace*{1mm}\noindent
\textbf{RQ3 (Performance).}
We assessed the effect of the adapter pattern~\cite{gof} to the size of software generated by \approach\ and the software after completing the migration 
and found that they are on average 40\% and 90\%, respectively, larger than the software given as input. 
For the environmental monitoring application, for instance, the Propeller software is 1,506 bytes compared to 2,156 and 2,922 bytes for the generated software and that from the completed migration, respectively. 
Since \approach\ generates placeholder methods for each Propeller-specific library method, a constant factor increase is expected. 
In the worst case, each line of the source software will make use of a Propeller library method, requiring three generated lines of code in the output software (method declaration; TODO directive; closing bracket).
More lines may be added for generated port commands and target platform include directives.
Without considering manual additions, the generated software has an upper bound $3L$ where $L$ is the number of lines in the source software (cf. Listing~\ref{lst:migratedCode}).
Furthermore, the size of the migrated software depends on the mappings between library methods for the source and target platforms. 
We found that achieving the same functionality using Arduino libraries requires more lines of code since component setup is generally more explicit than when using Propeller libraries (e.g. consider the servo and infrared transmitter examples discussed earlier). 
As such, software migrated directly to the target platform may be larger anyway without the use of the \approach\ approach.

We did not identify any perceptible difference in performance when running the robotic applications on both source and target platforms. Other than platform differences, the only potential performance reduction could come from using additional method calls from the generated methods.
However, compilers optimized for speed can inline functions to prevent this from occurring. 
Thus, the difference is likely to be negligible unless there is a cache miss. Other than this, performance differences are more likely to depend on the efficiency of the corresponding platform libraries and the hardware specifications for the platforms themselves.

\vspace*{1mm}\noindent
\textbf{RQ4 (Generality).} 
While our evaluation targets software migration from Propeller C to \CC,
\approach\ also supports migration of other combinations of 
procedural (C) and object-oriented (\CC) programming paradigms.
This entails that the source platform software (including platform-specific libraries) is in C/\CC\ and the migration is 
C$\rightarrow$C, C$\rightarrow$\CC\ or \CC$\rightarrow$\CC. 
Specialising the adapter pattern to migrate from \CC\ to C, albeit feasible, needs advanced constructs (e.g., functions in C structures) that outweigh the benefits provided, and thus, manual migration or re-development may be 
preferred. 

Beyond robotic systems, applying \approach\ to other domains such as cyber-physical systems and Internet-of-Things requires small modifications.
In particular, 
the analysis and library recommendation steps (Sections~\ref{ssec:approachStep1} and~\ref{ssec:approachStep3}) require no changes subject to the source and target platform software combinations discussed above and the availability of a set of possible candidate libraries.
Determining hardware components and communication ports for other Fritzing specifications is natively supported (Section~\ref{ssec:approachStep2}), while supporting inputs form similar tools (e.g., KiCad, Eagle) requires small adaptations to the text-to-model transformation. 
Finally, the template for software generation (Section~\ref{ssec:approachStep4}) should be adapted to match the structure expected by the target platform.

\subsection{Threats to Validity}
We mitigate \textbf{construct validity} threats that could be due to simplifications in the experimental setup
by using robotic applications with complexity and size similar to applications in real-world scenarios.
We also employ 
hardware components widely-used in applications  available on open-source robotic repositories (e.g., \url{https://create.arduino.cc/projecthub}).

We limit \textbf{internal validity} threats that could lead to bias in the process adopted for validating \approach\ by  assessing the correctness of each \approach\ step individually through comparing the generated against the expected outputs.  
We reduce further these threats by manually completing the migration for both robotic applications and confirming their successful migration from a Propeller Activity board to an Arduino Uno.


We address \textbf{external validity} threats that could affect the generalisation of the \approach\ tool-supported instance by using open-source and extensible software components widely-adopted in MDE.
The source platform software and target platform hardware analysis steps are applicable to any robotic software written in C/C++ that could be analysed using Eclipse CDT~\cite{eclipseCDT} and to any hardware configuration provided as a Fritzing specification, respectively. 
The library recommendation step is generalisable to other applications while the TDIDF relevance score heuristic could be easily replaced with more sophisticated techniques. 
Some modifications might be needed for software generation for other target platforms. For example, the generated software currently conforms to the Arduino structure of \code{setup} and \code{loop} methods. 
Thus, our findings are not conclusive for all types of robotic software migration activities, and more experiments are needed to confirm the generality and scalability of the \approach\ approach and tool.

%% file: sections/s6-relatedWork.tex

\section{Related Work}\label{sec:relatedWork}\
One of the earliest projects focusing on model-based migration and modernisation was \cite{fleurey2007model}, which explored the use of reverse engineering, model transformation and model migration techniques to migrate a mainframe application to a more modern J2EE application. The novelty in this work was the development of a model-based process, which extracted a model from (legacy) source code. This model was then transformed into a platform-independent model (consisting of representations of static data structures, actions, application navigation logic, and user interfaces). The platform-independent model would then be transformed into a platform-specific (UML) model and thereafter to code. Some user intervention is typically required to ensure that the platform-independent model accurately captures all important information in the code. The approach, therefore, is not fully automated but nevertheless demonstrates improvements in the cost/effort in carrying out migration. There are also challenges in terms of supporting testing of the migrated application: test cases are not in general migrated with the application.

An OMG initiative to support standardization efforts in software migration was embodied in the \textit{Architecture Driven Modernization (ADM)} initiative~\cite{ADM}. This led to the development of a set of standards to support reverse engineering, transformation, and migration activities. Of particular note is the \textit{Knowledge Discovery Metamodel (KDM)} standard, which is a metamodel for representing the key knowledge pertaining to software assets in an enterprise system. It thus allows representation of the structure and behaviour of existing software systems at different levels of granularity (via its container concept). It provides facilities for representing software's operating environment, events and state transition behaviour, UI features, and persistent data. KDM is a representation standard, not an implementation, and does not provide mechanisms to support different software migration tasks.

The MoDisco framework~\cite{bruneliere2014modisco} is an Eclipse project that has produced tools for modernising existing software systems, taking a pattern-based approach and using software models. It implements metamodels to describe existing systems, pattern-based discoverers for querying existing systems and gathering information needed to populate migrated models. The tools in MoDisco are themselves generic and support migration of documentation or code, while aiming to support different quality assurance processes. Thus, new discoverers can be built for new programming languages. It can also make use of KDM as a representation for the results of discoverers. 

In \cite{ellison2018evaluating}, the authors demonstrate an approach, with a supporting toolchain, for software migration of the data layer of data-intensive applications to cloud infrastructure, based on the use of model transformations and the KDM. The approach is two-stage, and accurately estimates the migration cost, migration duration and cloud running costs of relational databases. The first stage obtains workload and structure models of the database to be migrated from database logs and the database schema. The second stage performs a discrete-event simulation using these models to obtain the cost and duration estimates. 
The approach focuses on cloud migration specifically, though it uses generic standards (e.g., KDM and other metamodels).

In \cite{Gerasimou2018}, the authors demonstrate a purely code-based approach to software migration, exploiting Eclipse-based parsing and code generation tools. The approach aims to facilitate API, programming language, and hardware platform migration by analyzing source code, validating the source code to ensure that migration is feasible (e.g., by identifying code fragments that will require complex patterns to be imposed), refactoring the code base, and then generating target code. The approach was applied to several large code bases, and was shown to be useful to support profiling and hot-spot analysis. i.e., in identifying parts of the source application that may be challenging and expensive to migrate.

%% file: sections/s7-conclusion.tex

\vspace*{-2mm}
\section{Conclusion}
\label{sec:conclusion}
We presented \approach, a model-driven approach to software migration of robotic applications between different hardware platforms. \approach~uses the software deployed on the source platform and a description of the architecture for the target platform to generate software that can run on the target platform and indicate areas that require manual adaptation by engineers. \approach\ has been evaluated for the migration of two robotic applications from a Propeller Activity Board to an Arduino Uno. 
We found that \approach~can generate the correct artefacts for the software migration between hardware platforms and that the migrated applications are functionally equivalent with the original ones, although they are larger in terms of lines of code than the original because of using the adapter pattern. 
We plan to enhance \approach\ with support for validating hardware architecture specifications using constraint languages and mapping inference techniques~\cite{Robillard2013TSE}, and explore how to take into account the non-functional properties of the system.
Finally, we plan to investigate possible \approach\ extensions to support other programming language paradigms and assess its applicability to more complex and industrial-level robotic applications.


%% file: main.bbl

\begin{thebibliography}{35}


\ifx \showCODEN    \undefined \def \showCODEN     #1{\unskip}     \fi
\ifx \showDOI      \undefined \def \showDOI       #1{#1}\fi
\ifx \showISBNx    \undefined \def \showISBNx     #1{\unskip}     \fi
\ifx \showISBNxiii \undefined \def \showISBNxiii  #1{\unskip}     \fi
\ifx \showISSN     \undefined \def \showISSN      #1{\unskip}     \fi
\ifx \showLCCN     \undefined \def \showLCCN      #1{\unskip}     \fi
\ifx \shownote     \undefined \def \shownote      #1{#1}          \fi
\ifx \showarticletitle \undefined \def \showarticletitle #1{#1}   \fi
\ifx \showURL      \undefined \def \showURL       {\relax}        \fi
\providecommand\bibfield[2]{#2}
\providecommand\bibinfo[2]{#2}
\providecommand\natexlab[1]{#1}
\providecommand\showeprint[2][]{arXiv:#2}

\bibitem[\protect\citeauthoryear{??}{IEC}{2007}]%
        {IEC}
 \bibinfo{year}{2007}\natexlab{}.
\newblock \bibinfo{title}{{IEC 62402:2007 Obsolescence management. Application
  guide}}.
\newblock
\newblock


\bibitem[\protect\citeauthoryear{??}{JSP}{2007}]%
        {JSPP886}
 \bibinfo{year}{2007}\natexlab{}.
\newblock \bibinfo{title}{{JSPP 886, Volume 7, Part 8.13: Obsolescence
  management}}.
\newblock
\newblock


\bibitem[\protect\citeauthoryear{Alelyani, Michel, Yang, Wade, Verma, and
  T{\"o}rngren}{Alelyani et~al\mbox{.}}{2019}]%
        {alelyani2019literature}
\bibfield{author}{\bibinfo{person}{Turki Alelyani}, \bibinfo{person}{Ronald
  Michel}, \bibinfo{person}{Ye Yang}, \bibinfo{person}{Jon Wade},
  \bibinfo{person}{Dinesh Verma}, {and} \bibinfo{person}{Martin T{\"o}rngren}.}
  \bibinfo{year}{2019}\natexlab{}.
\newblock \showarticletitle{A literature review on obsolescence management in
  COTS-centric cyber physical systems}.
\newblock \bibinfo{journal}{\emph{Procedia computer science}}
  \bibinfo{volume}{153} (\bibinfo{year}{2019}), \bibinfo{pages}{135--145}.
\newblock


\bibitem[\protect\citeauthoryear{Bartolomei, Czarnecki, L{\"a}mmel, and Van
  Der~Storm}{Bartolomei et~al\mbox{.}}{2009}]%
        {bartolomei2009study}
\bibfield{author}{\bibinfo{person}{Thiago~Tonelli Bartolomei},
  \bibinfo{person}{Krzysztof Czarnecki}, \bibinfo{person}{Ralf L{\"a}mmel},
  {and} \bibinfo{person}{Tijs Van Der~Storm}.} \bibinfo{year}{2009}\natexlab{}.
\newblock \showarticletitle{Study of an API migration for two XML APIs}. In
  \bibinfo{booktitle}{\emph{International Conference on Software Language
  Engineering}}. Springer, \bibinfo{pages}{42--61}.
\newblock


\bibitem[\protect\citeauthoryear{Bergerman, Billingsley, Reid, and van
  Henten}{Bergerman et~al\mbox{.}}{2016}]%
        {bergerman2016robotics}
\bibfield{author}{\bibinfo{person}{Marcel Bergerman}, \bibinfo{person}{John
  Billingsley}, \bibinfo{person}{John Reid}, {and} \bibinfo{person}{Eldert van
  Henten}.} \bibinfo{year}{2016}\natexlab{}.
\newblock \showarticletitle{Robotics in agriculture and forestry}.
\newblock In \bibinfo{booktitle}{\emph{Springer handbook of robotics}}.
  \bibinfo{publisher}{Springer}, \bibinfo{pages}{1463--1492}.
\newblock


\bibitem[\protect\citeauthoryear{Bruneliere, Cabot, Dup{\'e}, and
  Madiot}{Bruneliere et~al\mbox{.}}{2014}]%
        {bruneliere2014modisco}
\bibfield{author}{\bibinfo{person}{Hugo Bruneliere}, \bibinfo{person}{Jordi
  Cabot}, \bibinfo{person}{Gr{\'e}goire Dup{\'e}}, {and}
  \bibinfo{person}{Fr{\'e}d{\'e}ric Madiot}.} \bibinfo{year}{2014}\natexlab{}.
\newblock \showarticletitle{Modisco: A model driven reverse engineering
  framework}.
\newblock \bibinfo{journal}{\emph{Information and Software Technology}}
  \bibinfo{volume}{56}, \bibinfo{number}{8} (\bibinfo{year}{2014}),
  \bibinfo{pages}{1012--1032}.
\newblock


\bibitem[\protect\citeauthoryear{Clavreul, Barais, and
  J{\'e}z{\'e}quel}{Clavreul et~al\mbox{.}}{2010}]%
        {clavreul2010integrating}
\bibfield{author}{\bibinfo{person}{Mickael Clavreul}, \bibinfo{person}{Olivier
  Barais}, {and} \bibinfo{person}{Jean-Marc J{\'e}z{\'e}quel}.}
  \bibinfo{year}{2010}\natexlab{}.
\newblock \showarticletitle{Integrating legacy systems with mde}. In
  \bibinfo{booktitle}{\emph{32nd ACM/IEEE International Conference on Software
  Engineering-Volume 2}}. ACM, \bibinfo{pages}{69--78}.
\newblock


\bibitem[\protect\citeauthoryear{Dig and Johnson}{Dig and Johnson}{2005}]%
        {Dig2005ICSM}
\bibfield{author}{\bibinfo{person}{Danny Dig} {and} \bibinfo{person}{Ralph
  Johnson}.} \bibinfo{year}{2005}\natexlab{}.
\newblock \showarticletitle{The Role of Refactorings in {API} Evolution}. In
  \bibinfo{booktitle}{\emph{21st IEEE International Conference on Software
  Maintenance (ICSM'05)}}. \bibinfo{pages}{389--398}.
\newblock


\bibitem[\protect\citeauthoryear{{Eclipse Foundation}}{{Eclipse
  Foundation}}{2010}]%
        {eclipseCDT}
\bibfield{author}{\bibinfo{person}{{Eclipse Foundation}}.}
  \bibinfo{year}{2010}\natexlab{}.
\newblock \bibinfo{title}{{Eclipse CDT}}.
\newblock \bibinfo{howpublished}{\url{https://www.eclipse.org/cdt}}.
\newblock
\newblock
\shownote{Accessed: 30-03-2019.}


\bibitem[\protect\citeauthoryear{Ellison, Calinescu, and Paige}{Ellison
  et~al\mbox{.}}{2018}]%
        {ellison2018evaluating}
\bibfield{author}{\bibinfo{person}{Martyn Ellison}, \bibinfo{person}{Radu
  Calinescu}, {and} \bibinfo{person}{Richard~F Paige}.}
  \bibinfo{year}{2018}\natexlab{}.
\newblock \showarticletitle{Evaluating cloud database migration options using
  workload models}.
\newblock \bibinfo{journal}{\emph{Journal of Cloud Computing}}
  \bibinfo{volume}{7}, \bibinfo{number}{1} (\bibinfo{year}{2018}),
  \bibinfo{pages}{6}.
\newblock


\bibitem[\protect\citeauthoryear{Farinelli, Zanotto, Pagello,
  et~al\mbox{.}}{Farinelli et~al\mbox{.}}{2017}]%
        {farinelli2017advanced}
\bibfield{author}{\bibinfo{person}{Alessandro Farinelli},
  \bibinfo{person}{Elena Zanotto}, \bibinfo{person}{Enrico Pagello},
  {et~al\mbox{.}}} \bibinfo{year}{2017}\natexlab{}.
\newblock \showarticletitle{Advanced approaches for multi-robot coordination in
  logistic scenarios}.
\newblock \bibinfo{journal}{\emph{Robotics and Autonomous Systems}}
  \bibinfo{volume}{90} (\bibinfo{year}{2017}), \bibinfo{pages}{34--44}.
\newblock


\bibitem[\protect\citeauthoryear{Fleurey, Breton, Baudry, Nicolas, and
  J{\'e}z{\'e}quel}{Fleurey et~al\mbox{.}}{2007}]%
        {fleurey2007model}
\bibfield{author}{\bibinfo{person}{Franck Fleurey}, \bibinfo{person}{Erwan
  Breton}, \bibinfo{person}{Benoit Baudry}, \bibinfo{person}{Alain Nicolas},
  {and} \bibinfo{person}{Jean-Marc J{\'e}z{\'e}quel}.}
  \bibinfo{year}{2007}\natexlab{}.
\newblock \showarticletitle{Model-driven engineering for software migration in
  a large industrial context}. In \bibinfo{booktitle}{\emph{International
  Conference on Model Driven Engineering Languages and Systems}}. Springer,
  \bibinfo{pages}{482--497}.
\newblock


\bibitem[\protect\citeauthoryear{Gamma}{Gamma}{1995}]%
        {gof}
\bibfield{author}{\bibinfo{person}{Erich Gamma}.}
  \bibinfo{year}{1995}\natexlab{}.
\newblock \bibinfo{booktitle}{\emph{Design patterns: elements of reusable
  object-oriented software}}.
\newblock \bibinfo{publisher}{Pearson Education India}.
\newblock


\bibitem[\protect\citeauthoryear{Gerasimou, Kechagia, Kolovos, Paige, and
  Gousios}{Gerasimou et~al\mbox{.}}{2018}]%
        {Gerasimou2018}
\bibfield{author}{\bibinfo{person}{Simos Gerasimou}, \bibinfo{person}{Maria
  Kechagia}, \bibinfo{person}{Dimitris Kolovos}, \bibinfo{person}{Richard
  Paige}, {and} \bibinfo{person}{Georgios Gousios}.}
  \bibinfo{year}{2018}\natexlab{}.
\newblock \showarticletitle{On Software Modernisation Due to Library
  Obsolescence}. In \bibinfo{booktitle}{\emph{2nd International Workshop on API
  Usage and Evolution}} \emph{(\bibinfo{series}{WAPI '18})}.
\newblock


\bibitem[\protect\citeauthoryear{Gerasimou and Kolovos}{Gerasimou and
  Kolovos}{2018}]%
        {cdtDriver}
\bibfield{author}{\bibinfo{person}{Simos Gerasimou} {and}
  \bibinfo{person}{Dimitris Kolovos}.} \bibinfo{year}{2018}\natexlab{}.
\newblock \bibinfo{title}{{Epsilon CDT Driver}}.
\newblock \bibinfo{howpublished}{\url{https://github.com/gerasimou/EMC-CDT}}.
\newblock
\newblock
\shownote{Accessed: 30-03-2019.}


\bibitem[\protect\citeauthoryear{Gerasimou, Kolovos, Paige, and
  Standish}{Gerasimou et~al\mbox{.}}{2017}]%
        {gerasimou2017technical}
\bibfield{author}{\bibinfo{person}{Simos Gerasimou}, \bibinfo{person}{Dimitris
  Kolovos}, \bibinfo{person}{Richard Paige}, {and} \bibinfo{person}{Michael
  Standish}.} \bibinfo{year}{2017}\natexlab{}.
\newblock \showarticletitle{Technical Obsolescence Management Strategies for
  Safety-Related Software for Airborne Systems}. In
  \bibinfo{booktitle}{\emph{Federation of International Conferences on Software
  Technologies: Applications and Foundations}}. Springer,
  \bibinfo{pages}{385--393}.
\newblock


\bibitem[\protect\citeauthoryear{Hawes, Burbridge, Jovan, Kunze, Lacerda,
  Mudrova, Young, Wyatt, Hebesberger, Kortner, et~al\mbox{.}}{Hawes
  et~al\mbox{.}}{2017}]%
        {hawes2017strands}
\bibfield{author}{\bibinfo{person}{Nick Hawes}, \bibinfo{person}{Christopher
  Burbridge}, \bibinfo{person}{Ferdian Jovan}, \bibinfo{person}{Lars Kunze},
  \bibinfo{person}{Bruno Lacerda}, \bibinfo{person}{Lenka Mudrova},
  \bibinfo{person}{Jay Young}, \bibinfo{person}{Jeremy Wyatt},
  \bibinfo{person}{Denise Hebesberger}, \bibinfo{person}{Tobias Kortner},
  {et~al\mbox{.}}} \bibinfo{year}{2017}\natexlab{}.
\newblock \showarticletitle{The {STRANDS} project: Long-term autonomy in
  everyday environments}.
\newblock \bibinfo{journal}{\emph{IEEE Robotics \& Automation Magazine}}
  \bibinfo{volume}{24}, \bibinfo{number}{3} (\bibinfo{year}{2017}),
  \bibinfo{pages}{146--156}.
\newblock


\bibitem[\protect\citeauthoryear{{International Federation of
  Robotics}}{{International Federation of Robotics}}{2016}]%
        {WRS}
\bibfield{author}{\bibinfo{person}{{International Federation of Robotics}}.}
  \bibinfo{year}{2016}\natexlab{}.
\newblock \showarticletitle{World Robotic Survey}.
\newblock  (\bibinfo{year}{2016}).
\newblock
\urldef\tempurl%
\url{https://ifr.org/news/world-robotics-survey-service-robots-are-conquering-the-world-/}
\showURL{%
\tempurl}


\bibitem[\protect\citeauthoryear{Izquierdo and Molina}{Izquierdo and
  Molina}{2014}]%
        {izquierdo2014extracting}
\bibfield{author}{\bibinfo{person}{Javier Luis~C{\'a}novas Izquierdo} {and}
  \bibinfo{person}{Jes{\'u}s~Garc{\'\i}a Molina}.}
  \bibinfo{year}{2014}\natexlab{}.
\newblock \showarticletitle{Extracting models from source code in software
  modernization}.
\newblock \bibinfo{journal}{\emph{Software \& Systems Modeling}}
  \bibinfo{volume}{13}, \bibinfo{number}{2} (\bibinfo{year}{2014}),
  \bibinfo{pages}{713--734}.
\newblock


\bibitem[\protect\citeauthoryear{Jennings, Wu, and Terpenny}{Jennings
  et~al\mbox{.}}{2016}]%
        {jennings2016forecasting}
\bibfield{author}{\bibinfo{person}{Connor Jennings}, \bibinfo{person}{Dazhong
  Wu}, {and} \bibinfo{person}{Janis Terpenny}.}
  \bibinfo{year}{2016}\natexlab{}.
\newblock \showarticletitle{Forecasting obsolescence risk and product life
  cycle with machine learning}.
\newblock \bibinfo{journal}{\emph{IEEE Transactions on Components, Packaging
  and Manufacturing Technology}} \bibinfo{volume}{6}, \bibinfo{number}{9}
  (\bibinfo{year}{2016}), \bibinfo{pages}{1428--1439}.
\newblock


\bibitem[\protect\citeauthoryear{Kn{\"o}rig, Wettach, and Cohen}{Kn{\"o}rig
  et~al\mbox{.}}{2009}]%
        {knorig2009fritzing}
\bibfield{author}{\bibinfo{person}{Andr{\'e} Kn{\"o}rig}, \bibinfo{person}{Reto
  Wettach}, {and} \bibinfo{person}{Jonathan Cohen}.}
  \bibinfo{year}{2009}\natexlab{}.
\newblock \showarticletitle{Fritzing: a tool for advancing electronic
  prototyping for designers}. In \bibinfo{booktitle}{\emph{Proceedings of the
  3rd International Conference on Tangible and Embedded Interaction}}. ACM,
  \bibinfo{pages}{351--358}.
\newblock


\bibitem[\protect\citeauthoryear{Kolovos, Paige, and Polack}{Kolovos
  et~al\mbox{.}}{2006}]%
        {kolovos2006epsilon}
\bibfield{author}{\bibinfo{person}{Dimitrios Kolovos},
  \bibinfo{person}{Richard~F Paige}, {and} \bibinfo{person}{Fiona~AC Polack}.}
  \bibinfo{year}{2006}\natexlab{}.
\newblock \showarticletitle{The Epsilon object language (EOL)}. In
  \bibinfo{booktitle}{\emph{European Conference on Model Driven
  Architecture-Foundations and Applications}}. Springer,
  \bibinfo{pages}{128--142}.
\newblock


\bibitem[\protect\citeauthoryear{Kolovos, Rose, Paige, and
  Garc{\i}a-Dom{\i}nguez}{Kolovos et~al\mbox{.}}{2010}]%
        {kolovos2010epsilon}
\bibfield{author}{\bibinfo{person}{Dimitrios Kolovos}, \bibinfo{person}{Louis
  Rose}, \bibinfo{person}{Richard Paige}, {and} \bibinfo{person}{Antonio
  Garc{\i}a-Dom{\i}nguez}.} \bibinfo{year}{2010}\natexlab{}.
\newblock \showarticletitle{The Epsilon book}.
\newblock \bibinfo{journal}{\emph{Structure}}  \bibinfo{volume}{178}
  (\bibinfo{year}{2010}), \bibinfo{pages}{1--10}.
\newblock


\bibitem[\protect\citeauthoryear{{Lloyd's Register Foundation}}{{Lloyd's
  Register Foundation}}{2016}]%
        {LRF}
\bibfield{author}{\bibinfo{person}{{Lloyd's Register Foundation}}.}
  \bibinfo{year}{2016}\natexlab{}.
\newblock \showarticletitle{Foresight review of robotics and autonomous
  systems}.
\newblock  (\bibinfo{year}{2016}).
\newblock
\urldef\tempurl%
\url{https://www.lrfoundation.org.uk/en/news/foresight-review-of-robotics-and-autonomous-systems}
\showURL{%
\tempurl}


\bibitem[\protect\citeauthoryear{{Object Management Group, Inc}}{{Object
  Management Group, Inc}}{2012}]%
        {ADM}
\bibfield{author}{\bibinfo{person}{{Object Management Group, Inc}}.}
  \bibinfo{year}{2012}\natexlab{}.
\newblock \showarticletitle{Architecture-Driven Modernization Task Force.}
\newblock  (\bibinfo{year}{2012}).
\newblock
\urldef\tempurl%
\url{http://adm.omg.org/}
\showURL{%
\tempurl}


\bibitem[\protect\citeauthoryear{Raibulet, Fontana, and Zanoni}{Raibulet
  et~al\mbox{.}}{2017}]%
        {raibulet2017model}
\bibfield{author}{\bibinfo{person}{Claudia Raibulet},
  \bibinfo{person}{Francesca~Arcelli Fontana}, {and} \bibinfo{person}{Marco
  Zanoni}.} \bibinfo{year}{2017}\natexlab{}.
\newblock \showarticletitle{Model-driven reverse engineering approaches: A
  systematic literature review}.
\newblock \bibinfo{journal}{\emph{IEEE Access}}  \bibinfo{volume}{5}
  (\bibinfo{year}{2017}), \bibinfo{pages}{14516--14542}.
\newblock


\bibitem[\protect\citeauthoryear{Rajagopal, Erkoyuncu, and Roy}{Rajagopal
  et~al\mbox{.}}{2015}]%
        {rajagopal2015impact}
\bibfield{author}{\bibinfo{person}{S Rajagopal}, \bibinfo{person}{JA
  Erkoyuncu}, {and} \bibinfo{person}{R Roy}.} \bibinfo{year}{2015}\natexlab{}.
\newblock \showarticletitle{Impact of software obsolescence in defence
  manufacturing sectors}.
\newblock \bibinfo{journal}{\emph{Procedia CIRP}}  \bibinfo{volume}{28}
  (\bibinfo{year}{2015}), \bibinfo{pages}{197--201}.
\newblock


\bibitem[\protect\citeauthoryear{Robillard, Bodden, Kawrykow, Mezini, and
  Ratchford}{Robillard et~al\mbox{.}}{2013}]%
        {Robillard2013TSE}
\bibfield{author}{\bibinfo{person}{Martin~P. Robillard}, \bibinfo{person}{Eric
  Bodden}, \bibinfo{person}{David Kawrykow}, \bibinfo{person}{Mira Mezini},
  {and} \bibinfo{person}{Tristan Ratchford}.} \bibinfo{year}{2013}\natexlab{}.
\newblock \showarticletitle{Automated {API} Property Inference Techniques}.
\newblock \bibinfo{journal}{\emph{IEEE Transactions on Software Engineering}}
  \bibinfo{volume}{39}, \bibinfo{number}{5} (\bibinfo{year}{2013}),
  \bibinfo{pages}{613--637}.
\newblock


\bibitem[\protect\citeauthoryear{Rose, Paige, Kolovos, and Polack}{Rose
  et~al\mbox{.}}{2008}]%
        {rose2008epsilon}
\bibfield{author}{\bibinfo{person}{Louis~M Rose}, \bibinfo{person}{Richard~F
  Paige}, \bibinfo{person}{Dimitrios~S Kolovos}, {and}
  \bibinfo{person}{Fiona~AC Polack}.} \bibinfo{year}{2008}\natexlab{}.
\newblock \showarticletitle{The Epsilon generation language}. In
  \bibinfo{booktitle}{\emph{European Conference on Model Driven
  Architecture-Foundations and Applications}}. Springer,
  \bibinfo{pages}{1--16}.
\newblock


\bibitem[\protect\citeauthoryear{Salton and McGill}{Salton and McGill}{1986}]%
        {salton1986introduction}
\bibfield{author}{\bibinfo{person}{Gerard Salton} {and}
  \bibinfo{person}{Michael~J McGill}.} \bibinfo{year}{1986}\natexlab{}.
\newblock \showarticletitle{Introduction to modern information retrieval}.
\newblock  (\bibinfo{year}{1986}).
\newblock


\bibitem[\protect\citeauthoryear{Sandborn and Myers}{Sandborn and
  Myers}{2008}]%
        {Sandborn2008:HPE}
\bibfield{author}{\bibinfo{person}{Peter Sandborn} {and}
  \bibinfo{person}{Jessica Myers}.} \bibinfo{year}{2008}\natexlab{}.
\newblock \bibinfo{booktitle}{\emph{Designing Engineering Systems for
  Sustainability}}.
\newblock \bibinfo{pages}{81--103}.
\newblock


\bibitem[\protect\citeauthoryear{Siciliano and Khatib}{Siciliano and
  Khatib}{2016}]%
        {siciliano2016springer}
\bibfield{author}{\bibinfo{person}{Bruno Siciliano} {and}
  \bibinfo{person}{Oussama Khatib}.} \bibinfo{year}{2016}\natexlab{}.
\newblock \bibinfo{booktitle}{\emph{Springer handbook of robotics}}.
\newblock \bibinfo{publisher}{Springer}.
\newblock


\bibitem[\protect\citeauthoryear{Tonelli~Bartolomei, Czarnecki, and
  Lammel}{Tonelli~Bartolomei et~al\mbox{.}}{2010}]%
        {tonelli2010swing}
\bibfield{author}{\bibinfo{person}{Thiago Tonelli~Bartolomei},
  \bibinfo{person}{Krzysztof Czarnecki}, {and} \bibinfo{person}{Ralf Lammel}.}
  \bibinfo{year}{2010}\natexlab{}.
\newblock \showarticletitle{Swing to SWT and back: Patterns for API migration
  by wrapping}. In \bibinfo{booktitle}{\emph{International Conference on
  Software Maintenance}}. IEEE, \bibinfo{pages}{1--10}.
\newblock


\bibitem[\protect\citeauthoryear{Wang, Wang, Mohammed, and Givehchi}{Wang
  et~al\mbox{.}}{2017}]%
        {wang2017ubiquitous}
\bibfield{author}{\bibinfo{person}{Xi~Vincent Wang}, \bibinfo{person}{Lihui
  Wang}, \bibinfo{person}{Abdullah Mohammed}, {and} \bibinfo{person}{Mohammad
  Givehchi}.} \bibinfo{year}{2017}\natexlab{}.
\newblock \showarticletitle{Ubiquitous manufacturing system based on Cloud: A
  robotics application}.
\newblock \bibinfo{journal}{\emph{Robotics and Computer-Integrated
  Manufacturing}}  \bibinfo{volume}{45} (\bibinfo{year}{2017}),
  \bibinfo{pages}{116--125}.
\newblock


\bibitem[\protect\citeauthoryear{Warren, Adams, and Molle}{Warren
  et~al\mbox{.}}{2011}]%
        {Warren2011}
\bibfield{author}{\bibinfo{person}{John-David Warren}, \bibinfo{person}{Josh
  Adams}, {and} \bibinfo{person}{Harald Molle}.}
  \bibinfo{year}{2011}\natexlab{}.
\newblock \bibinfo{booktitle}{\emph{Arduino for Robotics}}.
\newblock \bibinfo{publisher}{Apress}, \bibinfo{pages}{51--82}.
\newblock


\end{thebibliography}
